\documentclass[pra,showpacs]{revtex4}
\usepackage{graphicx}
\usepackage{subfigure}
\usepackage{caption2}
\usepackage{float}
\usepackage{amsmath}
\usepackage{amssymb}
\usepackage{color}

\begin{document}

\title{Spontaneous soliton symmetry breaking in two-dimensional coupled
Bose-Einstein condensates supported by optical lattices}
\author{ Arthur Gubeskys and Boris A. Malomed}
\affiliation{Department of Physical Electronics, School of Electrical Engineering,
Faculty of Engineering, Tel Aviv University, Tel Aviv 69978, Israel}

\begin{abstract}
Models of two-dimensional (2D) traps, with the double-well structure in the
third direction, for Bose-Einstein condensate (BEC) are introduced, with
attractive or repulsive interactions between atoms. The models are based on
systems of linearly coupled 2D Gross-Pitaevskii equations (GPEs), where the
coupling accounts for tunneling between the wells. Each well carries an
optical lattice (OL) (stable 2D solitons cannot exist without OLs). The
linear coupling splits each finite bandgap in the spectrum of the
single-component model into two \textit{subgaps}. The main subject of the
work is spontaneous symmetry breaking (SSB) in two-component 2D solitons and
localized vortices (SSB was not considered before in 2D settings). Using
variational approximation (VA) and numerical methods, we demonstrate that,
in the system with attraction or repulsion, SSB occurs in families of
symmetric or antisymmetric solitons (or vortices), respectively. The
corresponding bifurcation destabilizes the original solution branch and
gives rise to a stable branch of asymmetric solitons or vortices. The VA
provides for an accurate description of the emerging branch of asymmetric
solitons. In the model with attraction, all stable branches eventually
terminate due to the onset of collapse. Stable asymmetric solitons in higher
finite bandgaps, and vortices with a multiple topological charge are found
too. The models also give rise to first examples of \textit{embedded solitons%
} and \textit{embedded vortices }(the states located inside Bloch bands) in
two dimensions. In the linearly-coupled system with opposite signs of the
nonlinearity in the two cores, two distinct types of stable solitons and
vortices are found, dominated by either the self-attractive component or the
self-repulsive one. In the system with a mismatch between the two OLs, a
\textit{pseudo-bifurcation} is found: when the mismatch attains its largest
value ($\pi $), the bifurcation does not happen, as branches of different
solutions asymptotically approach each other but fail to merge.
\end{abstract}

\maketitle

\section{Introduction}

A great variety of dynamical patterns in Bose-Einstein condensates
(BECs) can be supported by optical lattices (OLs), i.e., spatially
periodic potentials induced by laser beams illuminating the BEC
\cite{Markus}. A well-known species of such patterns are gap
solitons in condensates composed of repulsively interacting atoms.
Gap solitons in BEC\ were predicted theoretically
\cite{GS}-\cite{Ostrovskaya2}, and then created in a
``cigar-shaped" (i.e., nearly one-dimensional, 1D) crossed-beam
optical trap, to which a longitudinal OL was added. In
self-attractive media, the OL potential makes it possible to trap
a usual soliton at a required position, and generates stable
multi-soliton complexes \cite{Wang}. In the case of a very deep
OL, the corresponding Gross-Pitaevskii equation (GPE) \cite{Pit}
reduces to a discrete nonlinear Schr\"{o}dinger (NLS) equation
\cite{Smerzi}, which supports well-known solitons, including
staggered ones \cite{Panos}.

Another topic which has drawn a great deal of interest, in terms of BEC \cite%
{double-well} and nonlinear optics \cite{double-well-optics}
alike, is \textit{spontaneous symmetry breaking} (SSB) in matter
waves or optical beams trapped in settings based on double-well
potentials (using the terminology widely adopted in optics
\cite{double-well-optics}, the medium corresponding to each
potential well is sometimes called a ``core" below). If two
parallel ``cigar-shaped" traps are strong enough, the system may
be described by a pair of one-dimensional GPEs with linear
coupling between them \cite{we}, hence the destabilization of
symmetric solitons and spontaneous transition to asymmetric ones
in the double cigar-shaped trap, filled with self-attractive
condensate, is described in the same way as the formation of
asymmetric solitons in the well-studied model of dual-core
nonlinear optical fibers \cite{dual-core} (in nonlinear optics,
the SSB of 1D solitons was also studied in dual-core models with
non-Kerr nonlinearities, such as quadratic \cite{chi2} and
cubic-quintic \cite{Lior}). On the other hand, if the dual-core
BEC\ waveguide is realized as a set of two parallel finite-width
potential troughs, separated by a finite buffer layer, which are
embedded in a 2D (horizontal) trap and tightly confined in the
transverse (vertical) direction, the SSB in the respective
double-core solitons can be examined in the framework of the full
2D GPE (i.e., considering the tunneling across the buffer layer
explicitly, rather than approximating it by the linear coupling
between the two 1D equations) \cite{Michal}. Both models predict a \textit{%
subcritical} symmetry-breaking bifurcation for solitons in the
self-attractive condensate [which means that branches of asymmetric
solutions emerge as unstable ones at the SSB point, and originally go
backward (for which reason the bifurcation is also called a backward one),
but then turn forward, stabilizing themselves at the turning point].

The objective of the present work is to find symmetric,
antisymmetric and asymmetric families of 2D BEC solitons, and
similar families of localized vortices (alias vortex solitons), in
the system built as a stacked pair of ``pancake-shaped" traps,
each carrying a two-dimensional OL. The traps are linearly coupled
by tunneling across the buffer layer separating them. This model
is a natural extension of a double-trap system recently studied in
the 1D setting \cite{we} (in optics, spontaneous symmetry breaking
in 1D multi-component gap solitons was studied in models of
dual-core \cite{Mak} and triple-core \cite{Arik} fiber Bragg
gratings). The generalization to the higher dimension is
significant, because 2D solitons are drastically different from
their 1D counterparts \cite{review}, especially in the case of the
self-attractive nonlinearity (the existence of the respective 2D
solitons is fundamentally restricted by the possibility of
collapse). Besides that, the new species of vortex solitons is
possible in 2D geometry \cite{review}. To the best of our
knowledge, SSB in 2D solitons or vortices has never been
considered before.

We will examine different physically relevant versions of the
model, with attractive or repulsive nonlinearity in each trap
(following Ref. \cite{we}, they are referred to as AA and RR
systems, i.e., ``attractive-attractive" and
``repulsive-repulsive"). In the AA system, the solitons originate
in the semi-infinite gap of the linear spectrum, while in the RR
system one may expect to find solitons in finite bandgaps induced
by the OL\ (we will consider the first two gaps, each of them
being split into two \textit{subgaps} under the action of the
linear coupling between the traps). In some cases (similar to what
was found in the 1D model \cite{we}), families of solitons and
vortices extend across Bloch bands separating the (sub)gaps, thus
becoming \textit{embedded solitons} \cite{embedded}. As a matter
of fact, the present paper reports the first examples of 2D
embedded solitons (and vortices).

A mixed (AR) system, with self-attraction in one trap and
self-repulsion in the parallel one, will be considered too. As is
known, the sign of the interaction between atoms can be reversed
by means of the Feshbach resonance imposed by external magnetic
field \cite{Feshbach}; accordingly, the signs may be made opposite
in the two traps by applying the magnetic field which is
inhomogeneous across the ``pancake" stack. Solitons in 1D
settings of the latter type, but without OLs, were elaborated in Refs. \cite%
{Valery}; a 2D generalization was considered too (with no OLs either) \cite%
{Valery2}, but it does not give rise to stable solitons.

In the normalized form, 2D models considered in this work are based on a
system of linearly coupled GPEs for the mean-field wave functions in the two
flat traps, $\psi (x,y,t)$ and $\phi (x,y,t)$,
\begin{subequations}
\begin{eqnarray}
i\psi _{t}+\nabla ^{2}\psi +\varepsilon \left[ \cos (2x)+\cos (2y)\right]
\psi +\lambda _{1}|\psi |^{2}\psi +\kappa \phi  &=&0,  \notag \\
&&  \label{the_model_2d} \\
i\phi _{t}+\nabla ^{2}\phi +\varepsilon \left[ \cos (2x+\Delta _{1})+\cos
(2y+\Delta _{2})\right] \phi +\lambda _{2}|\phi |^{2}\phi +\kappa \psi  &=&0,
\notag
\end{eqnarray}%
where $x$ and $y$ are planar coordinates, $\nabla ^{2}$ $\equiv \partial
^{2}/\partial x^{2}+\partial ^{2}/\partial y^{2}$, $\lambda _{1,2}=+1$ and $%
-1$ correspond to the attractive and repulsive nonlinearity in the
cores, real linear-coupling coefficient $\kappa $ is proportional
to the rate of tunneling across the potential barrier separating
the parallel traps (``cores"), and $\varepsilon $ is the strength
of the OLs, whose period is scaled to be $\pi $. We assume equal
depths of the OLs in both traps, while asymmetry between the
lattices may be introduced by
mismatches (phase shifts) in the $x$- and $y$-directions, $\Delta _{1}$ and $%
\Delta _{2}$. We will chiefly deal with the symmetric system, i.e., $\Delta
_{1}=\Delta _{2}=0$; nevertheless, effects produced by the mismatches will
be addressed too (in fact, we will consider the case of the largest possible
mismatch in the diagonal or horizontal direction, i.e., $\Delta _{1}=\Delta
_{2}=\pi $, or $\Delta _{1}=\pi $, $\Delta _{2}=0$, respectively). In
optical models, various effects of the mismatch between linearly-coupled
Bragg gratings on gap solitons supported by the gratings were studied in
Refs. \cite{Yossi}.

The 2D form of Eqs. (\ref{the_model_2d}) implies that the
``pancake"-shaped traps are tightly confined in the transverse
($Z$) direction, which allows one to reduce the underlying 3D GPE
to two dimensions, as performed, in the general form, in Refs.
\cite{3D-2D}. Together with normalizations of the OL period and
nonlinearity constants
adopted in Eqs. (\ref{the_model_2d}), this implies that variables $t$ and $%
\left( x,y\right) $ in Eqs. (\ref{the_model_2d}) are related to physical
time $T$ and coordinates $\left( X,Y\right) $ by
\end{subequations}
\begin{equation}
t\equiv \frac{\pi ^{2}\hslash }{2md^{2}}T,~\left( x,y\right) \equiv \frac{%
\pi }{d}\left( X,Y\right) ,  \label{scaling}
\end{equation}%
where $m$ is the atomic mass, and $d$ the OL period in physical units, while
the scaled wave functions are related to their counterparts, $\Psi $ and $%
\Phi $, in the 3D space as follows:
\begin{equation}
\left\{
\begin{array}{c}
\Psi (X,Y,Z,T) \\
\Phi (X,Y,Z,T)%
\end{array}%
\right\} =\frac{1}{2}\sqrt{\frac{\pi }{\left\vert a_{s}\right\vert d^{2}}}%
\left\{
\begin{array}{c}
\psi (x,t) \\
\phi (x,t)%
\end{array}%
\right\} \exp \left( -\frac{i}{2}\omega _{\perp }T-\frac{m\omega _{\perp }}{%
2\hbar }Z^{2}\right) .  \label{psi}
\end{equation}%
Here $\omega _{\perp }$ is the transverse trapping frequency, and $a_{s}$
the $s$-wave scattering length of atomic collisions ($a_{s}<0$ for
attractive interactions).\textrm{\ }Due to the normalizations, the lattice
strength is represented by $\varepsilon =2\mathcal{E}/\mathcal{E}_{\mathrm{%
rec}}$, where $\mathcal{E}_{\mathrm{rec}}=\left( \pi \hslash \right)
^{2}/\left( md^{2}\right) $\ is the lattice recoil energy, and $\mathcal{E}$
the depth of the periodic potential in physical units.

The AA and RR systems may be experimentally realized in $^{7}$Li \cite%
{Lithium} or $^{85}$Rb \cite{Cornish}, and $^{87}$Rb \cite{Markus}
condensates, respectively. In this work, the analysis is presented in the
range of $\kappa \sim 1$ in Eqs. (\ref{the_model_2d}). With regard to Eqs. (%
\ref{scaling}), the corresponding normalized tunnel-coupling time, $t_{%
\mathrm{coupl}}=\pi /(2\kappa )$, translates, for $d\sim 1$ $\mu $m, into
physical time $T_{\mathrm{coupl}}\sim 10$ $\mu $s and $100$ $\mu $s, for
lithium and rubidium, respectively. A prediction most essential to the
experiment is the number of atoms expected in stable solitons. Assuming an
experimentally relevant value of the trapping frequency -- for instance, $%
\omega _{\perp }\simeq 2\pi \times 100$ Hz for the rubidium condensate
(then, the corresponding transverse-confinement size is estimated to be $%
\sim 1$ $\mu $m, i.e., comparable to the OL period) -- and translating the
results reported below into physical units, we conclude that the
symmetry-breaking 2D solitons are built of $\sim 10^{3}$ atoms. In vortex
solitons, the number of atoms is larger, roughly, by a factor of $10$.

The paper is organized as follows. In Section II, we focus on relevant
mathematical formulations, which include the linear spectrum of the coupled
system, stationary equations for solitons, and the setting for the
variational approximation (VA) describing symmetric and asymmetric solitons
in the AA and RR systems. Systematic numerical results for solitons in
symmetric AA and RR systems (and comparison with predictions of the VA) are
reported in Section III, while Section IV reports numerical results for
vortex solitons, in the same systems. Findings for asymmetric systems (of
the AR type, as well as ones with mismatched OLs) are collected in Section
V. Section VI\ concludes the paper.

\section{Formulations: the linear spectrum, stationary equations, and
variational approximation}

\subsection{Linear spectra}

Before looking for soliton solutions, it is necessary to identify the
system's spectrum within the framework of the linearized equations. First,
we consider the symmetric system, with $\Delta _{1}=\Delta _{2}=0$, and look
for solutions with chemical potential $\mu $ as%
\begin{equation}
\left\{ \psi (x,y,t),\phi (x,y,t)\right\} =\left[ \alpha (x,y)\pm \beta (x,y)%
\right] e^{-i\mu t}.
\end{equation}%
Substituting this in the linearized version of Eqs. (\ref{the_model_2d}), we
arrive at a 2D eigenvalue problem based on the decoupled equations,
\begin{subequations}
\begin{eqnarray}
\nabla ^{2}\alpha (x,y)+\varepsilon \left[ \cos (2x)+\cos (2y)\right] \alpha
(x,y) &=&-(\mu +\kappa )\alpha (x,y), \\
\nabla ^{2}\beta (x,y)+\varepsilon \left[ \cos (2x)+\cos (2y)\right] \beta
(x,y) &=&-(\mu -\kappa )\beta (x,y).
\end{eqnarray}%
Each one of these equations is tantamount to the eigenvalue problem for the
single GPE with a 2D lattice and effective chemical potential $\mu ^{\prime
}=\mu \pm \kappa $. The spectrum generated by the linearization of the
single-component GPE is well known, see Refs. \cite{Ostrovskaya} and
references therein. Figure \ref{spectrum_fig} shows a typical example of the
spectrum of the 2D coupled system with the zero mismatch. We observe that
the linear coupling in Eqs. (\ref{the_model_2d}) splits the finite gaps of
the single-component model BEC into pairs of \textit{subgaps}, which is
similar to what was observed, under the action of the linear coupling, in
the 1D counterpart of the present system \cite{we}.

\begin{figure}[tbp]
\subfigure[]{\includegraphics[width=3in]{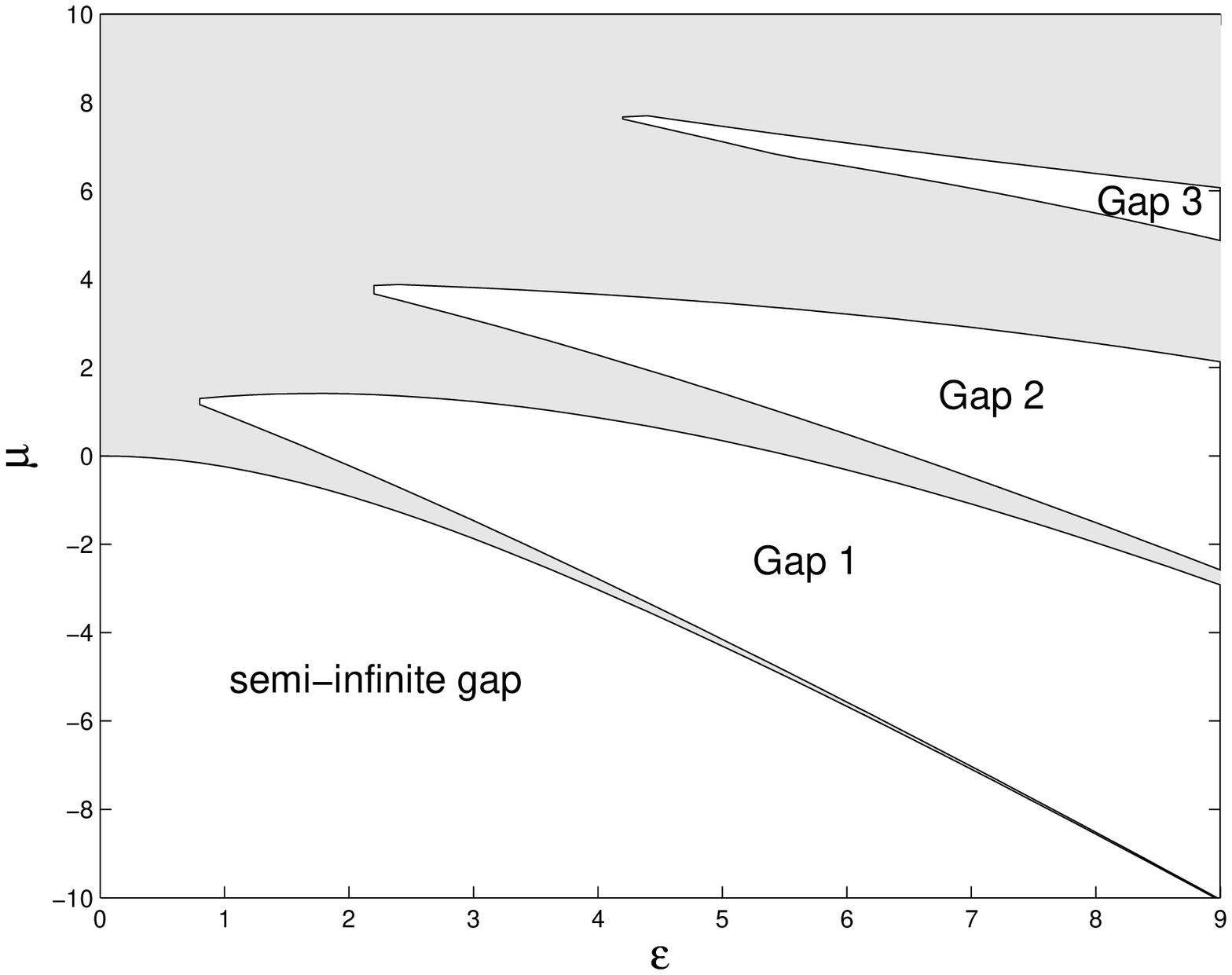}} \subfigure[]{%
\includegraphics[width=3in]{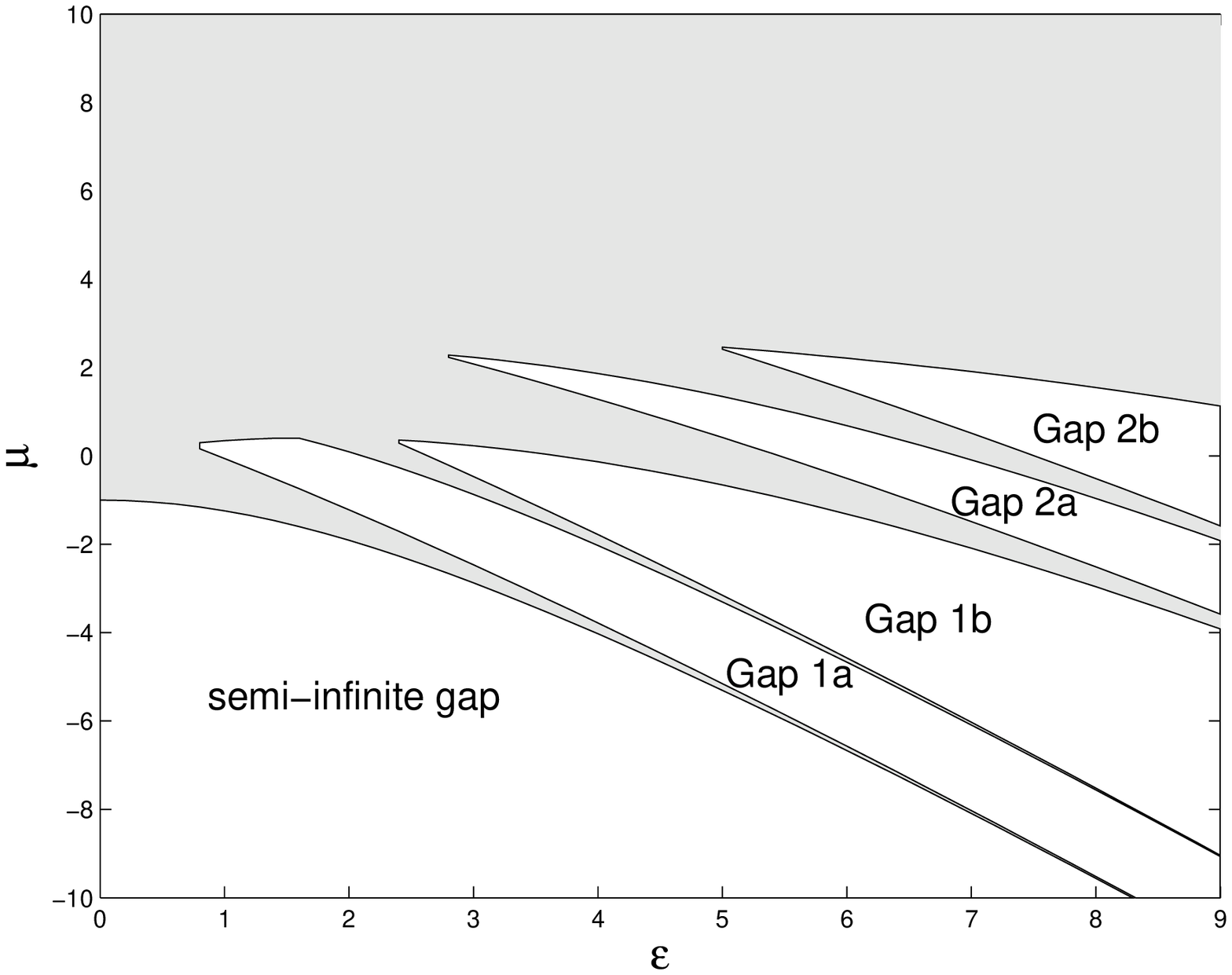}}
\caption{(Color online) The spectrum of the linearized 2D system with
aligned optical lattices ($\Delta _{1}=\Delta _{2}=0$). Throughout the
paper, all numerically obtained results are given for $\protect\varepsilon =8
$, which makes it possible to display the findings in a generic form. Here
and in figures following below, Bloch bands are shaded. (a) The uncoupled
system, $\protect\kappa =0$; (b) the coupled one, with $\protect\kappa =1$
(unless specified otherwise, all figures in the paper pertain to $\protect%
\kappa =1$).}
\label{spectrum_fig}
\end{figure}

To show in detail how the gaps split into subgaps, in Fig. \ref{extra} we
display the dependence of the chemical potential on vectorial quasi-momentum
$\left( k_{x},k_{y}\right) $ in the diagonal direction ($k_{x}=k_{y}\equiv k$%
), in the uncoupled system and in its coupled counterpart with $\kappa =1$,
both pertaining to $\varepsilon =8$. The new, nearly flat, branch of the $%
\mu (k)$ dependence, which appears in the latter case around $\mu =-7.5$,
explains the splitting of former gap 1 into subgaps 1a and 1b by a new very
narrow Bloch band, cf. Fig. \ref{spectrum_fig}. At other values of $%
\varepsilon $, the splitting mechanism is quite similar.

\begin{figure}[tbp]
\subfigure[]{\includegraphics[width=3in]{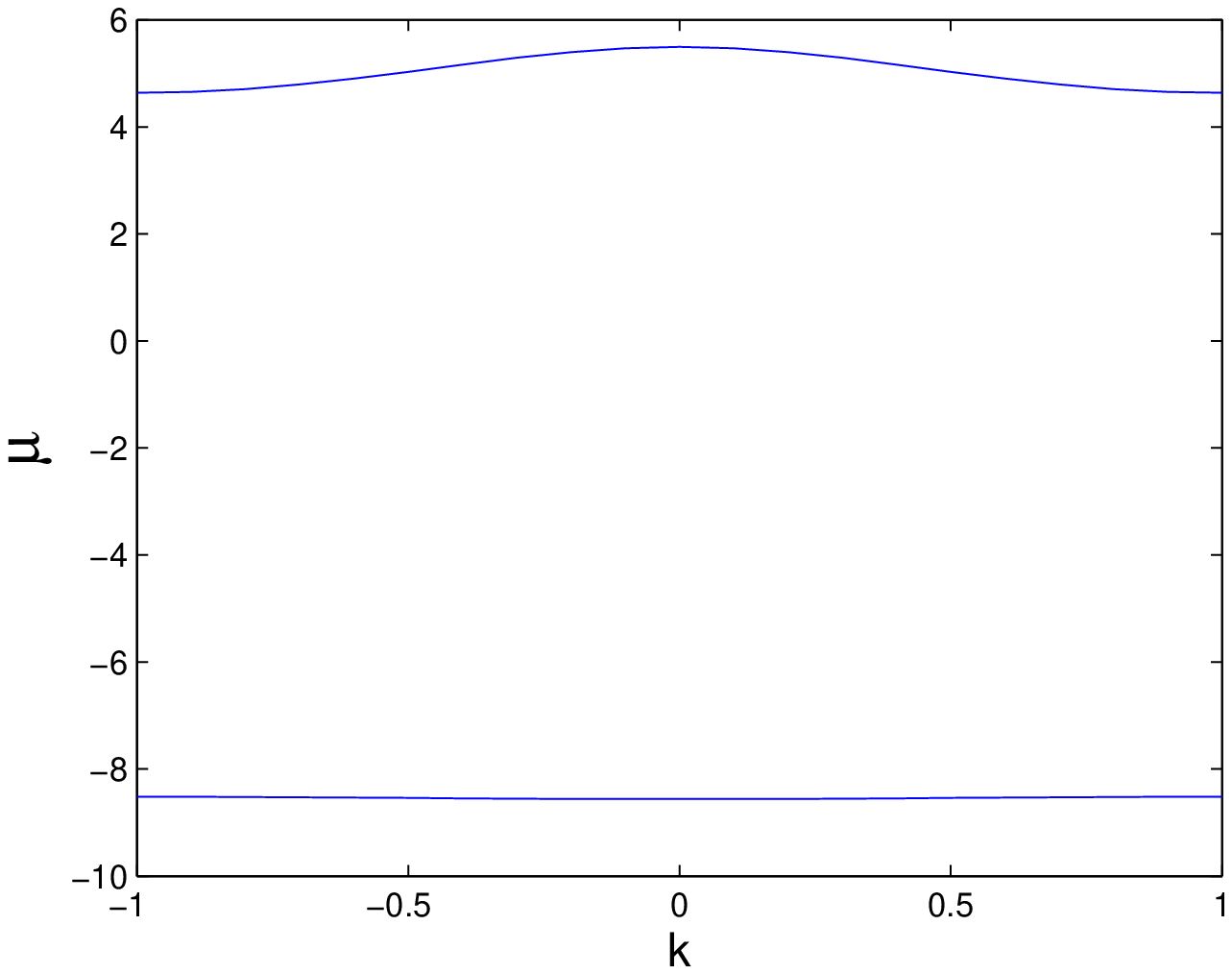}} \subfigure[]{%
\includegraphics[width=3in]{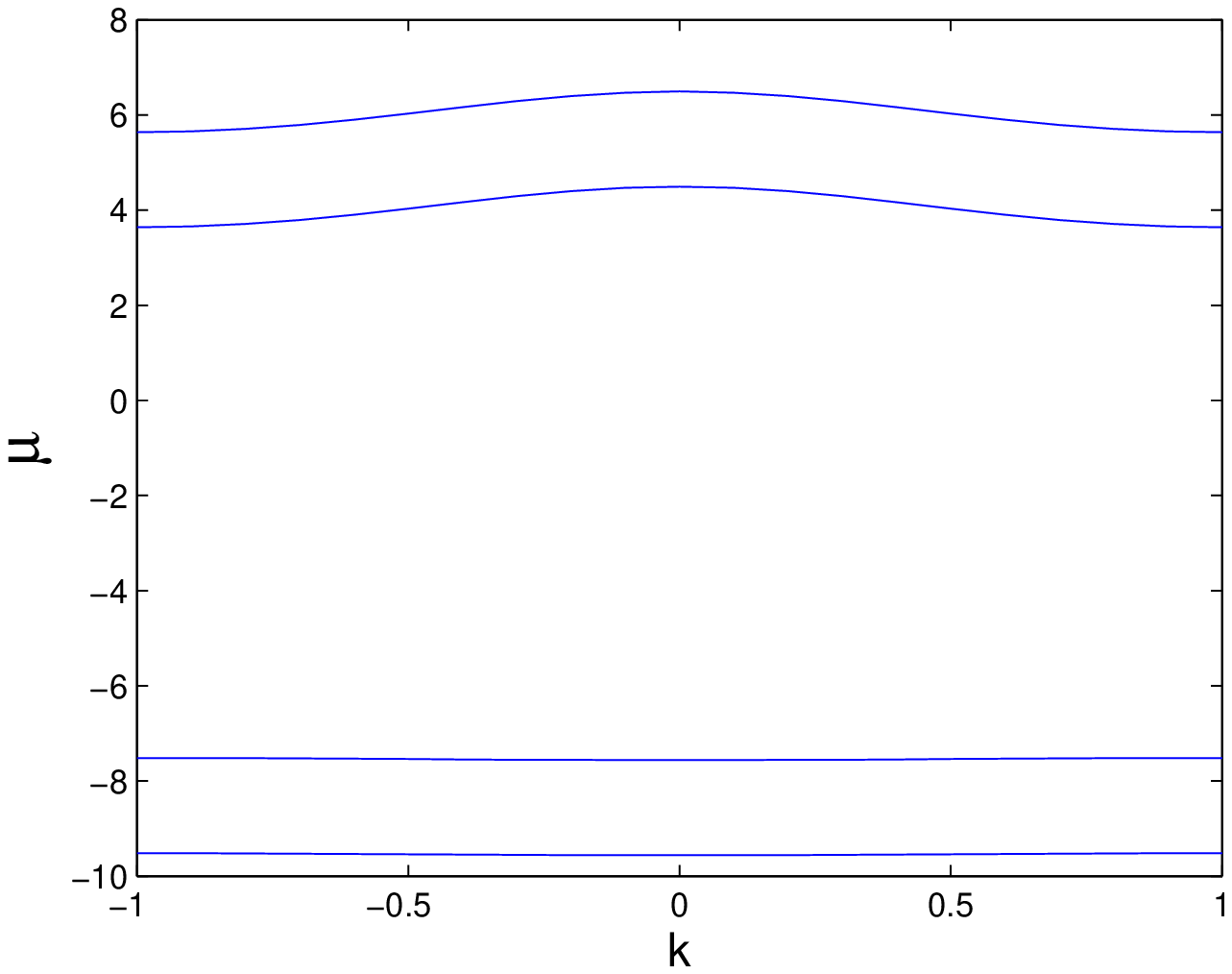}}
\caption{(Color online) The dependence of chemical potential $\protect\mu $
on quasi-momentum $k$ (in the diagonal direction, see text) in the uncoupled
system, $\protect\kappa =0$ (a), and in the coupled one, $\protect\kappa =1$ (b). In both cases, $\Delta _{1}=\Delta _{2}=0$ and $\varepsilon=8$. The new nearly flat branch of the dependence,
observed around $\protect\mu =-7.5$, explains the splitting of gap 1 into
subgaps 1a and 1b in Fig. \protect\ref{spectrum_fig}.}
\label{extra}
\end{figure}

Figure \ref{extra} and all others in this paper display numerical results
for OL strength $\varepsilon =8$ (a moderately deep lattice), which
corresponds to the most typical situation. As for coupling constant $\kappa $%
, the results are displayed for $\kappa =1$ (which also helps to presents
generic findings), unless a different value of $\kappa $ is indicated.

As shown in Fig. \ref{spectrum_delta}, we also examined the transformation
of the linear spectrum under the action of the diagonal mismatch between the
OLs in the two traps, assuming $\Delta _{1}=\Delta _{2}\equiv \Delta $. In
this case, the linearized equations cannot be decoupled; nevertheless, the
computation is straightforward, as we can use the separability of the linear
operator to construct eigenfunctions of the 2D model as products of
eigenfunctions of the corresponding 1D models \cite{we}. Figure \ref%
{spectrum_delta} demonstrates that, with the increase of $\Delta $, the
subgaps generated by the coupling-induced splitting tend to shrink, which
resembles the trend observed in the 1D model \cite{we}.

\begin{figure}[tbp]
\centering\includegraphics[width=3in]{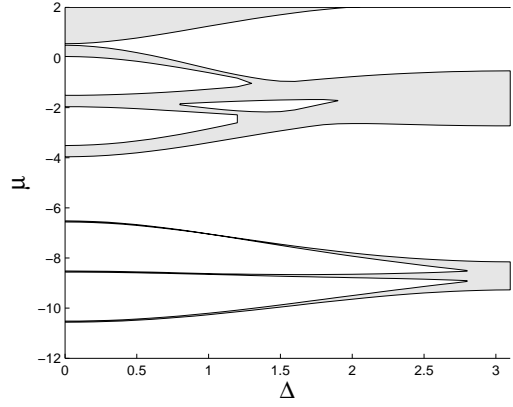}
\caption{(Color online) The change of the linear spectrum with the increase
of diagonal mismatch $\Delta $ between the lattices in the coupled traps,
for $\protect\kappa =2$.}
\label{spectrum_delta}
\end{figure}

\subsection{Stationary equations}

Stationary solutions to the full nonlinear system of Eqs. (\ref{the_model_2d}%
) are looked for as $\left\{ \psi ,\phi \right\} =\left\{
u(x,y),v(x,y)\right\} e^{-i\mu t}$, with real functions $u$ and $v$ to be
found from equations
\end{subequations}
\begin{subequations}
\begin{eqnarray}
\mu u+\nabla ^{2}u+\varepsilon \left[ \cos (2x)+\cos (2y)\right] u+\lambda
_{1}u^{3}+\kappa v &=&0,  \notag \\
&&  \label{stationary_2d} \\
\mu v+\nabla ^{2}v+\varepsilon \left[ \cos (2x+\Delta _{1})+\cos (2y+\Delta
_{2})\right] v+\lambda _{2}v^{3}+\kappa u &=&0.  \notag
\end{eqnarray}%
In the symmetric models, i.e., AA and RR ones ($\lambda _{1}=\lambda
_{2}=\pm 1$) with zero mismatch ($\Delta _{1}=\Delta _{2}=0$), symmetric ($%
u=v$) and antisymmetric ($u=-v$) solutions can be expressed in terms of a
stationary solution of the single-component GPE with chemical potential $\mu
$, to be denoted as $\hat{u}_{0}(x,y;\mu )$:
\end{subequations}
\begin{equation}
u(x,y;\mu )=\pm v(x,y;\mu )=\hat{u}_{0}(x,y;\mu \pm \kappa ).
\label{symm_sol_2d}
\end{equation}%
Similar to the situation in the 1D model \cite{we}, we conclude from Eq. (%
\ref{symm_sol_2d}) that, when the gap splitting occurs, the symmetric and
antisymmetric solitons will be moved to the lower and upper subgaps,
respectively. In some cases, they may end up in Bloch bands separating the
subgaps, thus becoming embedded solitons, examples of which are presented
below.

To find general asymmetric soliton solutions and SSB bifurcations linking
them to their symmetric and antisymmetric counterparts, we solved the full
system of stationary equations (\ref{stationary_2d}) numerically. The
stability of all solitons was examined by direct simulations of Eqs. (\ref%
{the_model_2d}). Soliton families are characterized (see below) by norms of
the two components and the total norm,
\begin{equation}
\left\{ N_{u},N_{v}\right\} =\int \int \left\{ u^{2}(x,y),v^{2}(x,y)\right\}
dxdy,~N\equiv N_{u}+N_{v},  \label{Norm}
\end{equation}%
For asymmetric solutions, we define the \textit{asymmetry ratio} \cite{we}
as
\begin{equation}
\Theta =\left\vert N_{u}-N_{v}\right\vert /\left( N_{u}+N_{v}\right) .
\label{theta}
\end{equation}%
The total norm, together with the energy, are dynamical invariants of Eqs. (%
\ref{the_model_2d}).

\subsection{Variational approximation for the symmetric system}

The symmetric version of Eqs. (\ref{stationary_2d}) ($\lambda _{1}=\lambda
_{2}\equiv \lambda ,~\Delta _{1}=\Delta _{2}=0$) can be derived from the
following Lagrangian:

\begin{eqnarray}
L &=&\int \int \left[ \mu \left( |u|^{2}+|v|^{2}\right) -\left( |\nabla
u|^{2}+|\nabla v|^{2}\right) \right. +\varepsilon \left[ \cos (2x)+\cos (2y)%
\right] \left( |u|^{2}+|v|^{2}\right)   \notag \\
&&\left. +\left( \lambda /2\right) \left( |u|^{4}+|v|^{4}\right) +\kappa
\left( u^{\ast }v+uv^{\ast }\right) \right] .  \label{L}
\end{eqnarray}%
A simple real isotropic Gaussian ansatz \cite{Baizakov}, with single width $W
$ but different amplitudes $A$ and $B$ pertaining to the two components, may
be adopted for the solitons:
\begin{equation}
\left( u,v\right) =\left( A,B\right) \exp \left( -\frac{x^{2}+y^{2}}{2W^{2}}%
\right) .  \label{ansatz}
\end{equation}%
The total norm of this expression is [see Eq. (\ref{Norm})]%
\begin{equation}
N=\pi W^{2}\left( A^{2}+B^{2}\right) .  \label{N}
\end{equation}

The substitution of ansatz (\ref{ansatz}) in Lagrangian (\ref{L}) yields the
corresponding \textit{effective Lagrangian},
\begin{eqnarray}
L_{\mathrm{eff}}/\pi  &=&\mu \left( A^{2}+B^{2}\right) W^{2}-\left(
A^{2}+B^{2}\right) +2\varepsilon \left( A^{2}+B^{2}\right) W^{2}e^{-W^{2}} \\
&&+\left( \lambda /4\right) \left( A^{4}+B^{4}\right) W^{2}+2\kappa ABW^{2}.
\end{eqnarray}%
Variational equations following from this Lagrangian can be conveniently
written as%
\begin{equation}
\partial L_{\mathrm{eff}}/\partial \left( A^{2}+B^{2}\right) =\partial L_{%
\mathrm{eff}}/\partial \left( AB\right) =\partial L_{\mathrm{eff}}/\partial
\left( W^{2}\right) =0.  \label{VA}
\end{equation}%
For symmetric and antisymmetric solitons, defined by $A=sB$ with $s=\pm 1$,
Eqs. (\ref{VA}) amount to a set of two equations,%
\begin{eqnarray}
N &\equiv &2\pi W^{2}A^{2}=8\pi \lambda \left( 1-2\varepsilon
W^{4}e^{-W^{2}}\right) ,  \label{Nsymm} \\
\mu +s\kappa  &=&-W^{-2}-2\varepsilon e^{-W^{2}}\left( 1-2W^{2}\right) .
\label{mu-symm}
\end{eqnarray}%
In the attractive model ($\lambda =+1$), Eqs. (\ref{Nsymm}) and (\ref%
{mu-symm}) are tantamount to those analyzed in the framework of the VA in
Refs. \cite{Baizakov} and \cite{GMM}. It was demonstrated that the norm of
the 2D soliton takes values in the following intervals:
\begin{eqnarray}
0 &<&N<N_{\max }^{(\mathrm{attr})}\equiv 8\pi ,~\mathrm{if~}\varepsilon
>\varepsilon _{\mathrm{cr}}\equiv e^{2}/8\approx \allowbreak 0.92;  \notag \\
&&  \label{+1Nmax} \\
N_{\min }^{(\mathrm{attr})} &\equiv &8\pi \left( 1-\varepsilon /\varepsilon
_{\mathrm{cr}}\right) <N<8\pi ,~\mathrm{if~}\varepsilon <\varepsilon _{%
\mathrm{cr}}.  \notag
\end{eqnarray}%
Note that $N_{\max }^{(\mathrm{attr})}=8\pi $ corresponds, in terms of the
VA \cite{Anderson}, to the norm of the \textit{Townes soliton}, i.e., the
localized solution of the 2D NLS equation in free space, which is unstable
against collapse \cite{Berge} (while the OL stabilizes 2D solitons \cite%
{Baizakov,Jianke}). In fact, the vanishing of $N_{\min }^{(\mathrm{attr})}$
at $\varepsilon >\varepsilon _{\mathrm{cr}}$ in Eqs. (\ref{+1Nmax}) is an
artifact of the VA \cite{Baizakov}, explained by the fact that Gaussian
ansatz (\ref{ansatz}) is not appropriate for multi-peaked solitons supported
by the strong OL.

In the repulsive model ($\lambda =-1$), variational equations (\ref{Nsymm})
and (\ref{mu-symm}) which, essentially, pertain to the single-component
setting, predict solitons only if the OL is strong enough \cite{GMM}.
Indeed, a straightforward consideration of Eq. (\ref{Nsymm}) demonstrates
that, in this case, solutions exist only for $\varepsilon >\varepsilon _{%
\mathrm{cr}}$, the norm of the solution family being limited to interval%
\begin{equation}
0<N<N_{\max }^{(\mathrm{rep})}\equiv 8\pi \left( \varepsilon /\varepsilon _{%
\mathrm{cr}}-1\right)   \label{-1Nmax}
\end{equation}%
[$\varepsilon _{\mathrm{cr}}$ is the same as in Eq. (\ref{+1Nmax})]. The
comparison of Eq. (\ref{-1Nmax}) with known numerical results for 2D gap
solitons \cite{Ostrovskaya,Sakaguchi} demonstrates that the nonexistence of
solitons in the repulsive model at $\varepsilon <\varepsilon _{\mathrm{cr}%
}^{(1)}$ is another artifact of the VA, signalling that gap solitons cannot
be approximated by simple Gaussian ansatz (\ref{ansatz}) for $\varepsilon
<\varepsilon _{\mathrm{cr}}$.

For asymmetric solitons, with $A^{2}\neq B^{2}$, Eqs. (\ref{VA}) can be cast
in the following form:%
\begin{eqnarray}
A^{2}+B^{2} &=&2\left[ \lambda \left( W^{-2}-2\varepsilon
W^{2}e^{-W^{2}}\right) +\sqrt{\left( W^{-2}-2\varepsilon
W^{2}e^{-W^{2}}\right) ^{2}+2\kappa ^{2}}\right] ,  \label{ABW} \\
AB &=&2\lambda \kappa ,  \label{AB} \\
\mu &=&-2\varepsilon e^{-W^{2}}\left( 1-W^{2}\right) -\lambda \sqrt{\left(
W^{-2}-2\varepsilon W^{2}e^{-W^{2}}\right) ^{2}+2\kappa ^{2}}.  \label{muW}
\end{eqnarray}%
With regard to Eq. (\ref{ABW}), norm (\ref{N}) of the asymmetric solitons
becomes%
\begin{equation}
N=2\pi \left[ \lambda \left( 1-2\varepsilon W^{4}e^{-W^{2}}\right) +\sqrt{%
\left( 1-2\varepsilon W^{4}e^{-W^{2}}\right) ^{2}+2\kappa ^{2}W^{4}}\right] ,
\label{NW}
\end{equation}%
Variational solutions for the asymmetric solitons are meaningful if they
satisfy an obvious condition, $A^{2}+B^{2}\geq 2\left\vert AB\right\vert $.
In fact, the asymmetric solitons bifurcate from symmetric or antisymmetric
ones [if, respectively, $\lambda =+1$ or $-1$, as follows from Eq. (\ref{AB}%
)] precisely at point $A^{2}+B^{2}=2\left\vert AB\right\vert $. Soliton
families predicted by the VA, i.e., obtained by numerical solution of Eqs. (%
\ref{ABW}) - (\ref{NW}), are represented by the corresponding dependences $%
N=N(\mu )$ in Figs. \ref{AA_solitons_family} and \ref{RR_solitons_family},
along with their counterparts found from the numerical solution of full
stationary equations (\ref{stationary_2d}).

\section{Numerical results: solitons}

\subsection{Symmetric attraction-attraction system}

\begin{figure}[tbp]
\subfigure[]{\includegraphics[width=3in]{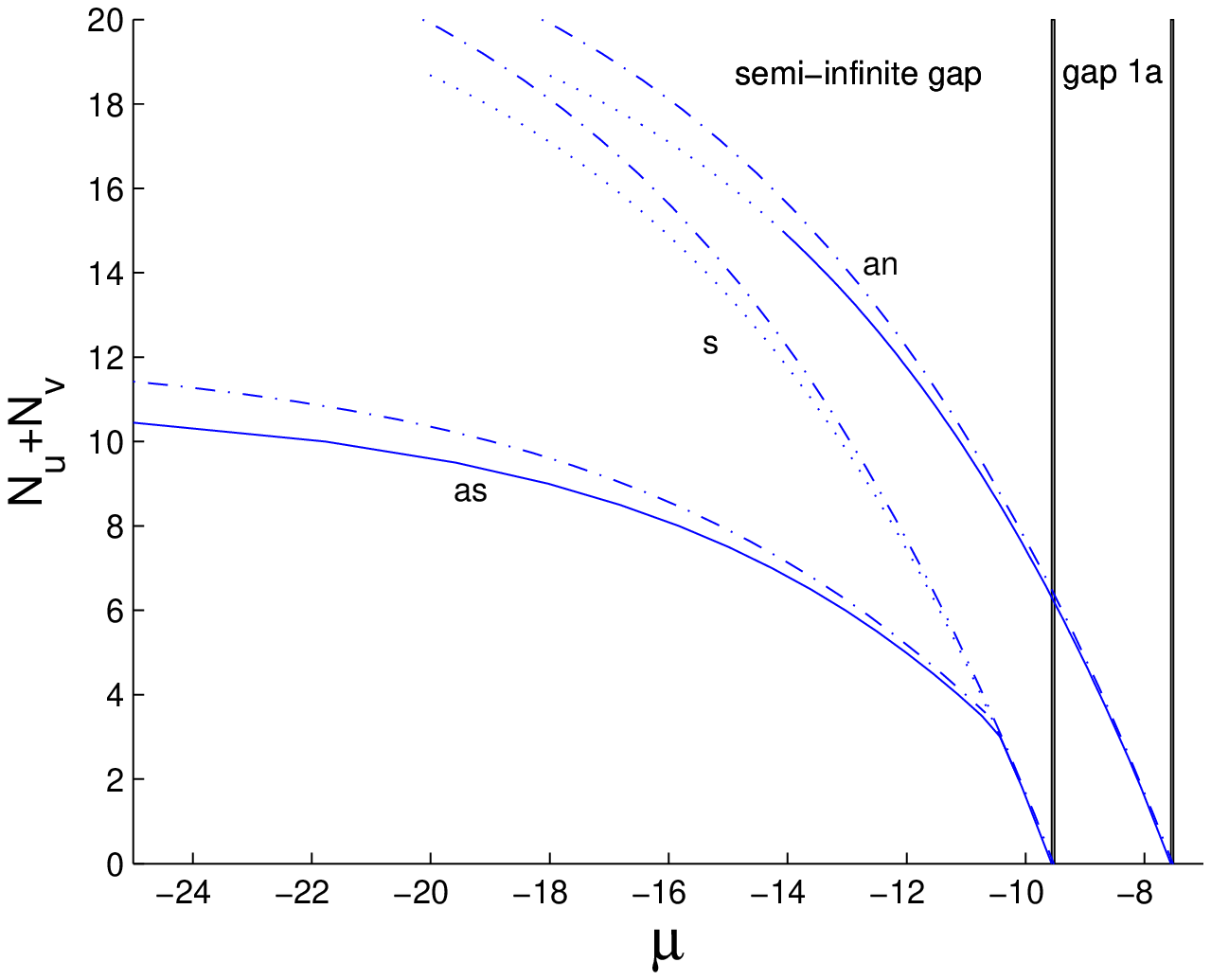}} \subfigure[]{%
\includegraphics[width=3in]{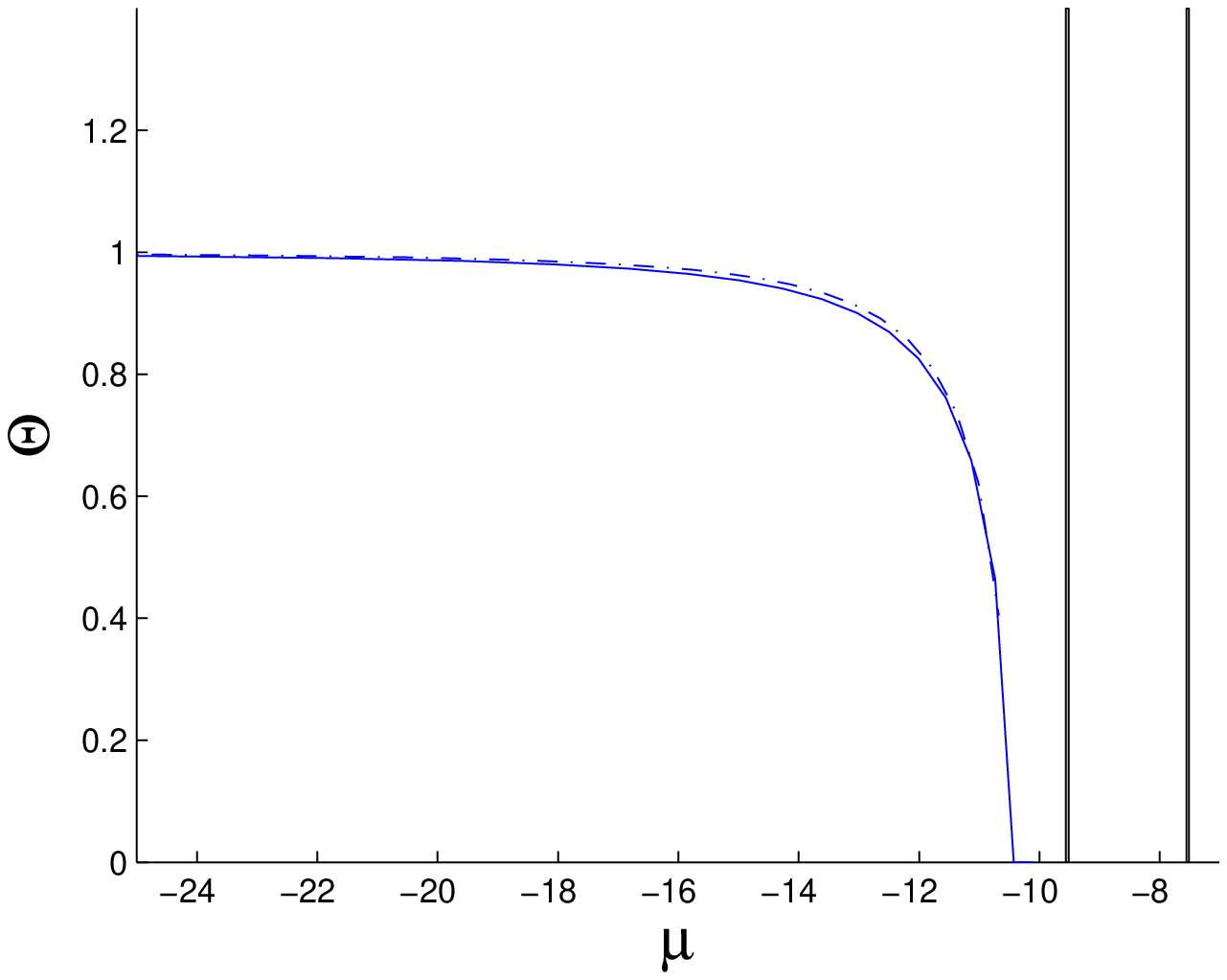}}
\caption{(Color online) Families of 2D solitons found in the symmetric
attraction-attraction model ($\protect\lambda _{1}=\protect\lambda _{2}=1$, $%
\Delta _{1}=\Delta _{2}=0$). Symmetric, antisymmetric and
asymmetric solutions are labeled by ``s", ``an", and ``as",
respectively. Dashed-dotted lines represent solutions (of all the
three types) produced by the variational approximation, while
solid and dotted lines depict, respectively, numerically found
stable and unstable solutions. (a) The soliton's total power
versus the chemical potential. (b) Ratio $\Theta $ [see Eq.
(\protect\ref{theta})] for the family of asymmetric solitons. The
same conventions for labelling solution families of different
types (variaitonal/numerical, stable/unstable,
symmetric/antisymmetric/asymmetric) are adopted in other figures.}
\label{AA_solitons_family}
\end{figure}

\begin{figure}[tbp]
\centering\includegraphics[width=3in]{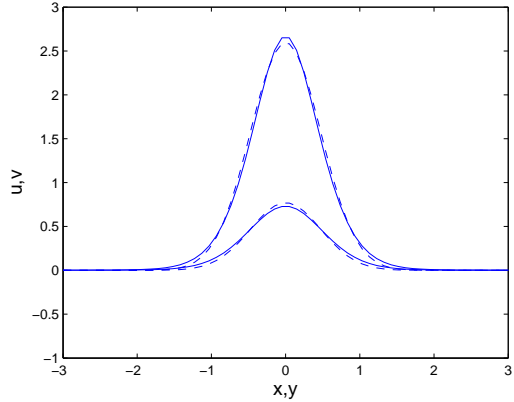}
\caption{(Color online) Comparison of profiles of the 2D soliton in the
attraction-attraction model, as found from the numerical solution and
predicted by the variational approximation (solid and dashed curves,
respectively). Shown are cross sections of the soliton along $x$- and $y$%
-axes, for $\protect\mu =-12$. Norms of the two components of the soliton
are $N_{u}\approx 4.5$ and $N_{v}\approx 0.4$.}
\label{va_vs_num_AA}
\end{figure}

In Fig. \ref{AA_solitons_family}, we display a generic example of families
of antisymmetric, symmetric and asymmetric stationary solitons in the AA
model with zero mismatch between the lattices ($\Delta =0$). The families
were found from systematic numerical solutions of Eqs. (\ref{stationary_2d}%
). Figure \ref{AA_solitons_family} also includes the solution families as
predicted, for the same case, by the VA, which demonstrates good agreement
between the variational and numerical results, for all the three types of
solitons (at small values of the norm, the variational branches are
indistinguishable from their numerical counterparts). A typical example of
comparison of the profiles of asymmetric solitons produced by the numerical
solution and VA is presented in Fig. \ref{va_vs_num_AA}.

We observe in Fig. \ref{AA_solitons_family}(a) that the symmetric-soliton
branch undergoes a \textit{supercritical} bifurcation at some critical value
of the norm, giving rise to the branch of asymmetric solitons, which is the
manifestation of the SSB in this setting (note that the bifurcation point is
very accurately predicted by the VA). Symmetric solitons are stable before
the bifurcation, and unstable afterwards, while asymmetric solitons emerge
as stable solutions [recall the stability of the solutions was identified by
direct simulations of Eqs. (\ref{the_model_2d})]. On the other hand, the
family of antisymmetric solitons never bifurcates, i.e., it never gets
destabilized through SSB. As a consequence of that, the system features
bistability, when the antisymmetric solutions are stable simultaneously with
either symmetric or asymmetric ones (below or above the bifurcation point,
respectively). It is worthy to note that all branches of the solutions
satisfy the Vakhitov-Kolokolov (VK) criterion, $dN/d\mu <0$, which is a
necessary stability condition \cite{VK,Berge}. It is also noteworthy that
the stable branch of antisymmetric solitons displayed in Fig. \ref%
{AA_solitons_family}(a) continues across the (narrow) Bloch band separating
the semi-infinite gap and subgap 1a. Inside the narrow band, this family
represents \emph{2D embedded solitons} (which are stable in direct
simulations). As far as we know, this is the first example of 2D embedded
solitons reported in any model.

\begin{figure}[tbp]
\subfigure[]{\includegraphics[width=3in]{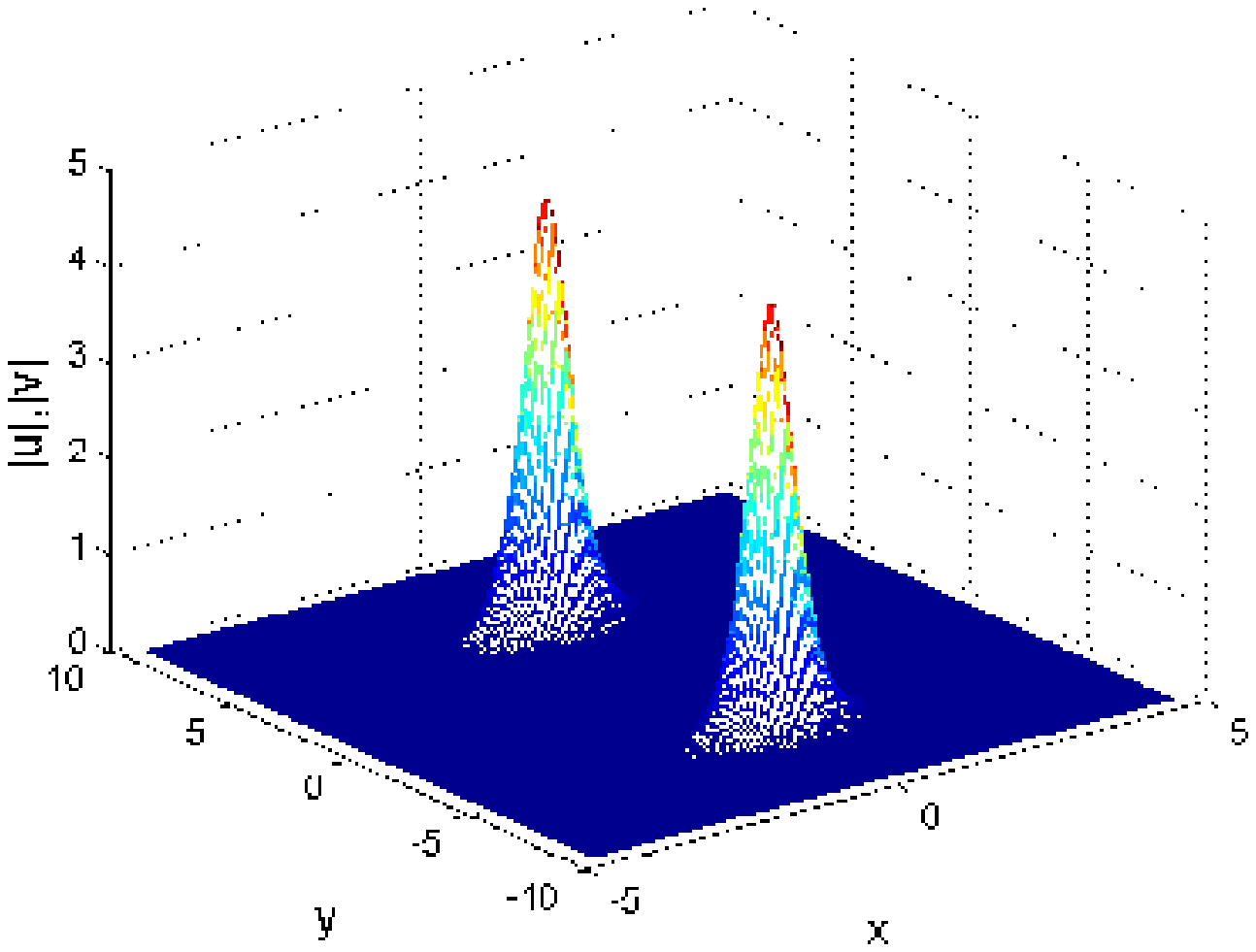}} \subfigure[]{%
\includegraphics[width=3in]{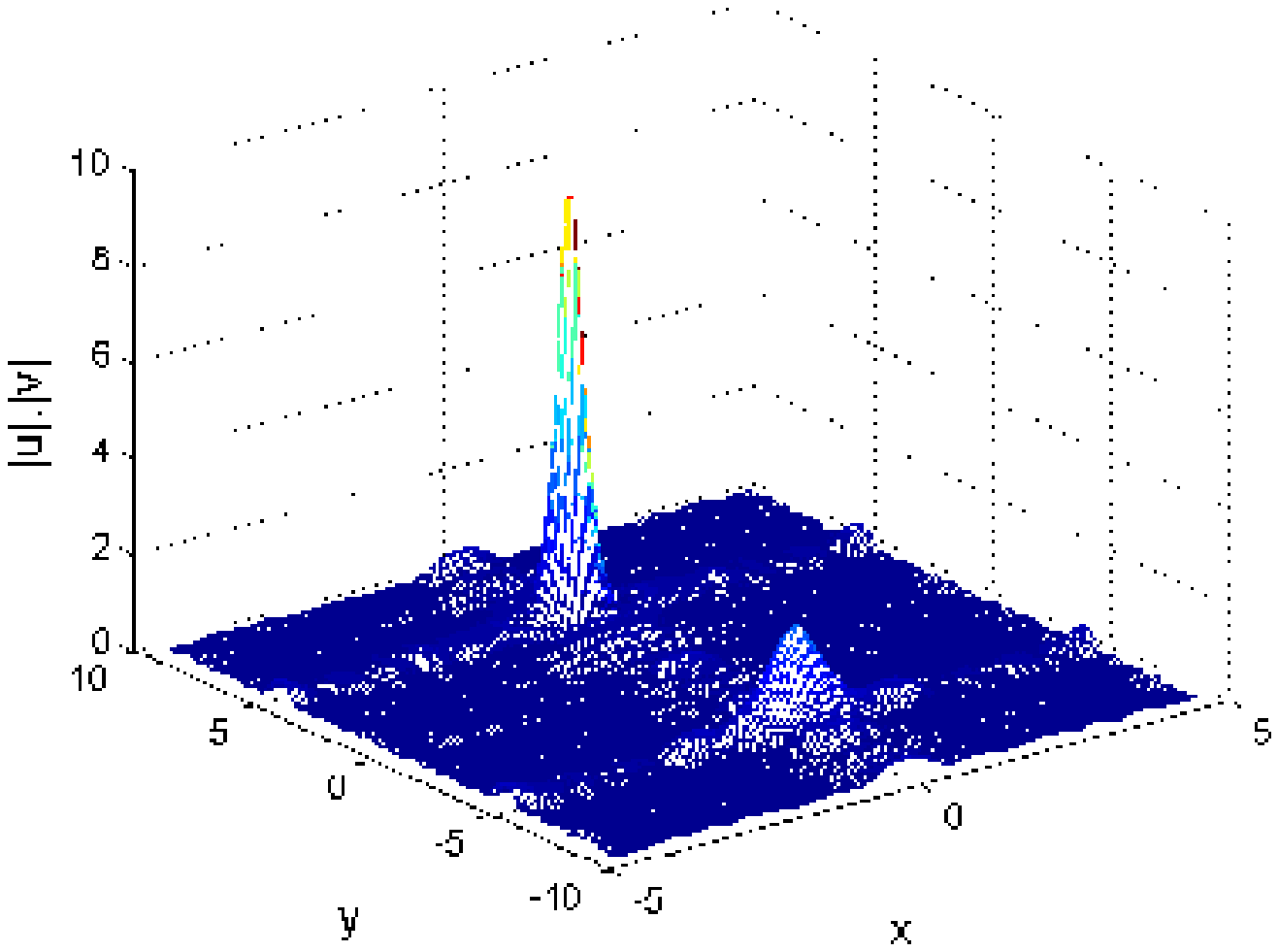}} \subfigure[]{%
\includegraphics[width=3in]{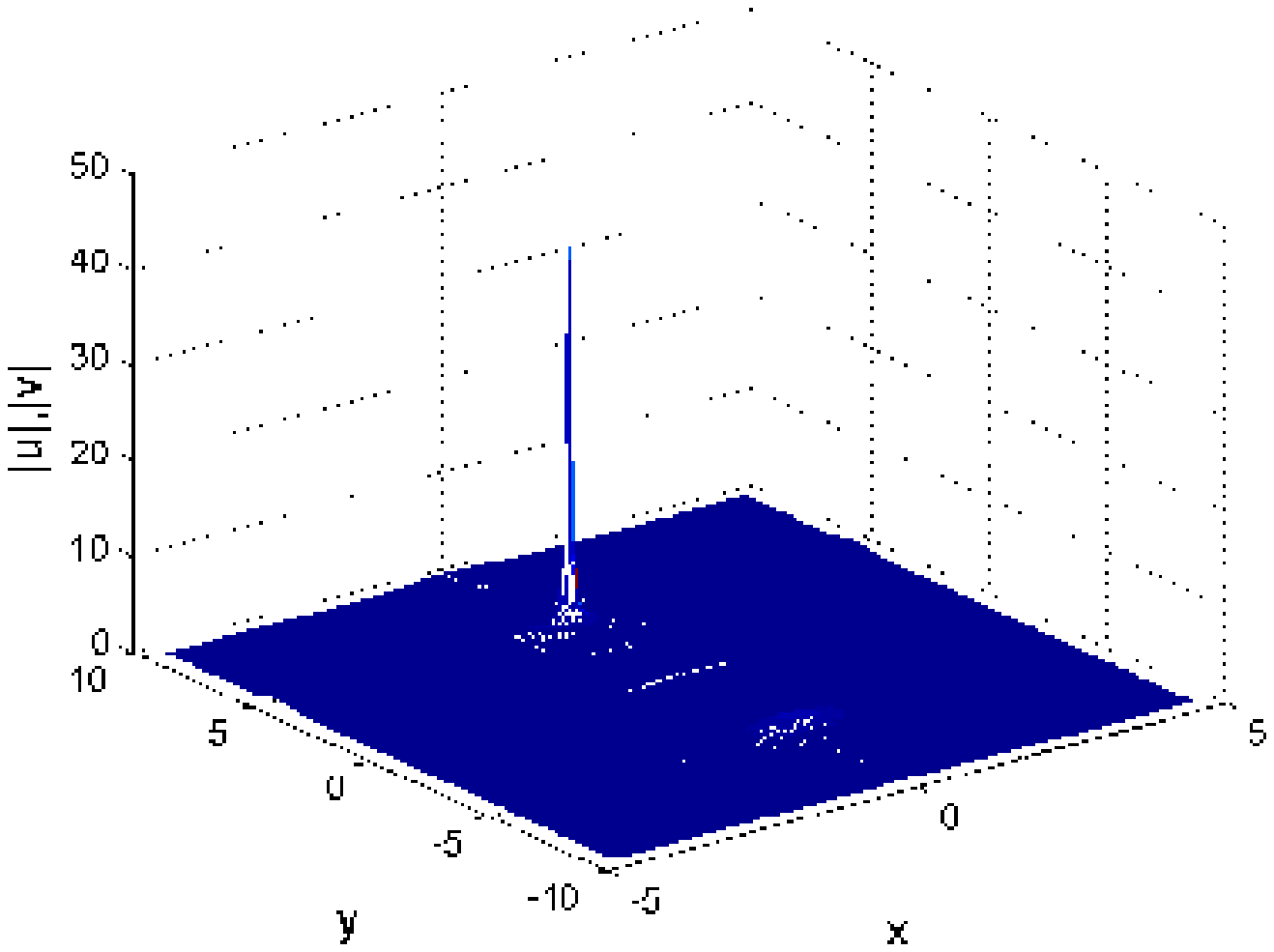}} \subfigure[]{%
\includegraphics[width=3in]{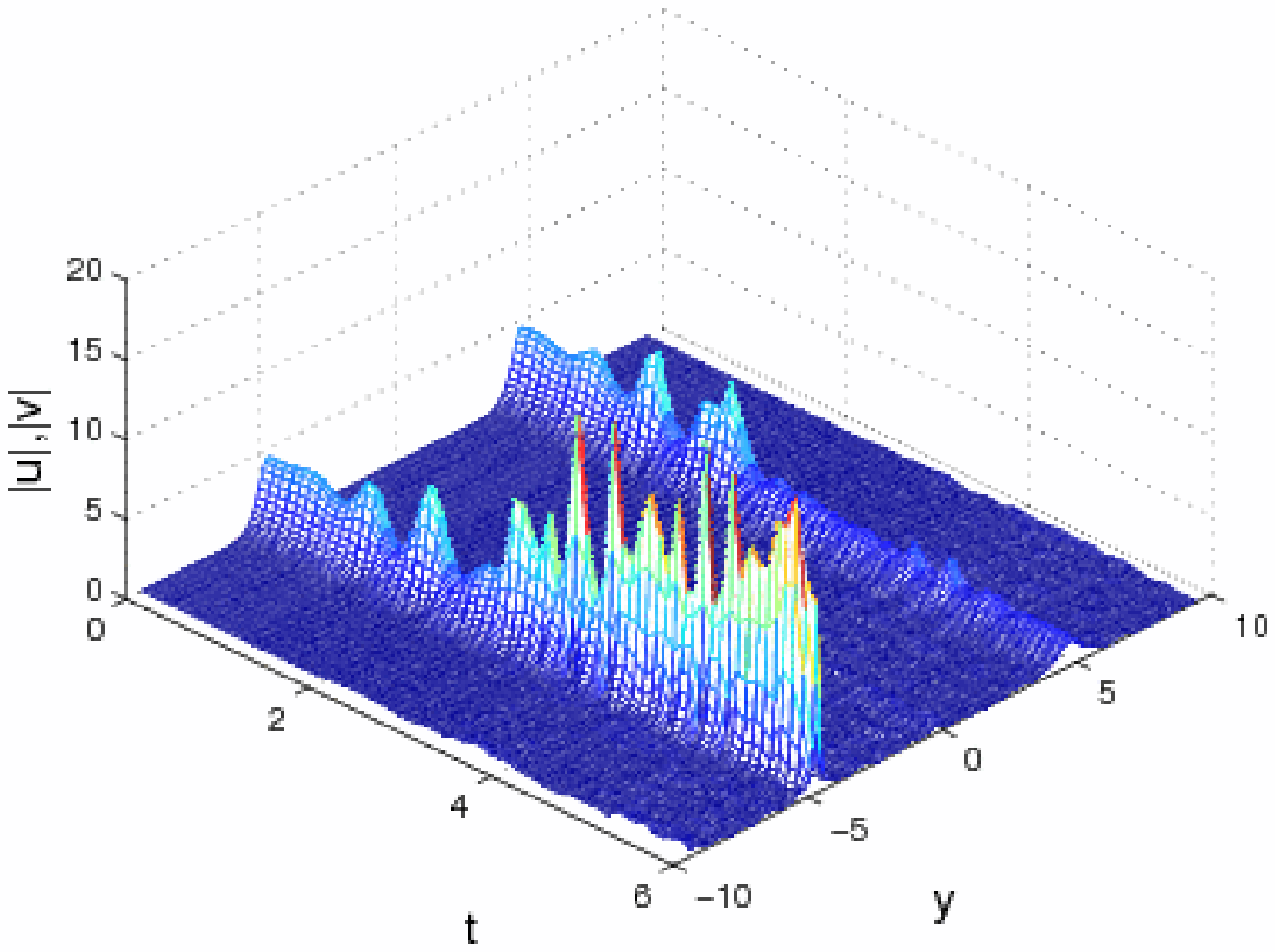}}
\caption{(Color online) Evolution of an antisymmetric 2D soliton (with $%
\protect\mu =-16$) in the attraction-attraction model, which is destabilized
by the collapse. (a) The initial state at $t=0$; (b) the formation of an
asymmetric state at $t=4$; (c) the collapse of the high-amplitude component
and decay of the low-amplitude one, at $t=6$; (d) the evolution of the
entire pattern in the cross section along $x=0$. Here and in similar figures
presented below, panels display the side-by-side juxtaposition of the two
components, $|u|$ and $|v|$.}
\label{antisymm_AA_collapse}
\end{figure}

The above results resemble findings recently reported in the 1D system \cite%
{we}. However, in the 2D case we observe an additional destabilization
mechanism in the AA system, through collapse of the soliton of any type,
when its norm becomes too large. In particular, the antisymmetric branch,
which would be totally stable in the 1D case, loses its stability in the 2D
system when the soliton's norm exceeds the corresponding collapse threshold.
Direct simulations demonstrate that, in this situation, the unstable
antisymmetric soliton is first transformed into a strongly asymmetric
structure; eventually, the high-amplitude component collapses, forming a
singularity, while its counterpart with the lower amplitude decays into
quasi-linear waves, as shown in Fig. \ref{antisymm_AA_collapse}.

The evolution of symmetric solitons destabilized by the SSB bifurcation is
not affected by the collapse. Similar to what was reported in the study of
the 1D model \cite{we}, the unstable symmetric solitons clearly tend to
rearrange themselves into stable asymmetric counterparts (examples are not
shown here, as they are not essentially different from what was observed in
the 1D system). The branch of the asymmetric solitons is also subject to the
collapse, but this happens on a remote portion of the $N(\mu )$ curve, which
could not be shown in Fig. \ref{AA_solitons_family}.

\subsection{Symmetric repulsion-repulsion system}

\begin{figure}[tbp]
\subfigure[]{\includegraphics[width=3in]{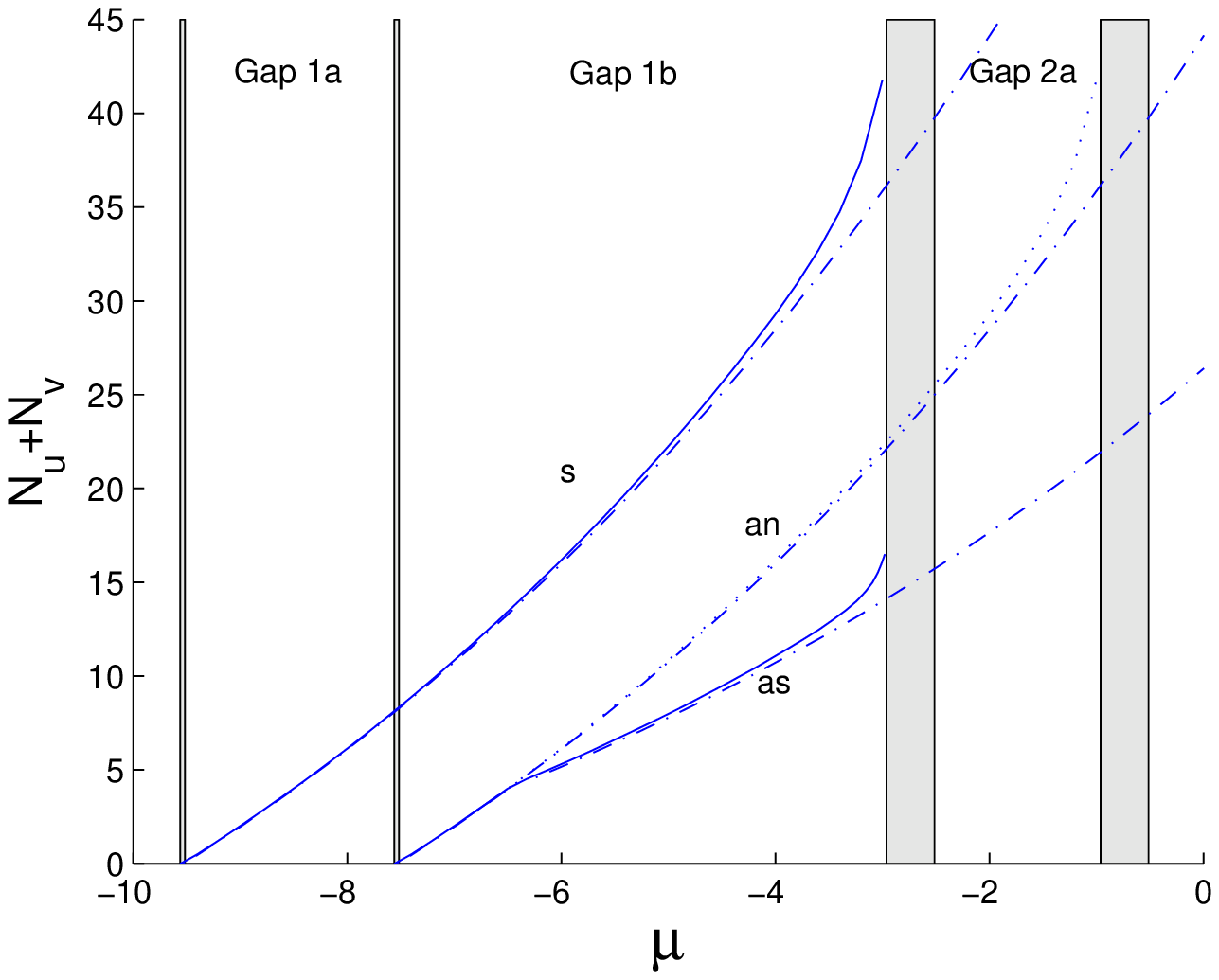}} \subfigure[]{%
\includegraphics[width=3in]{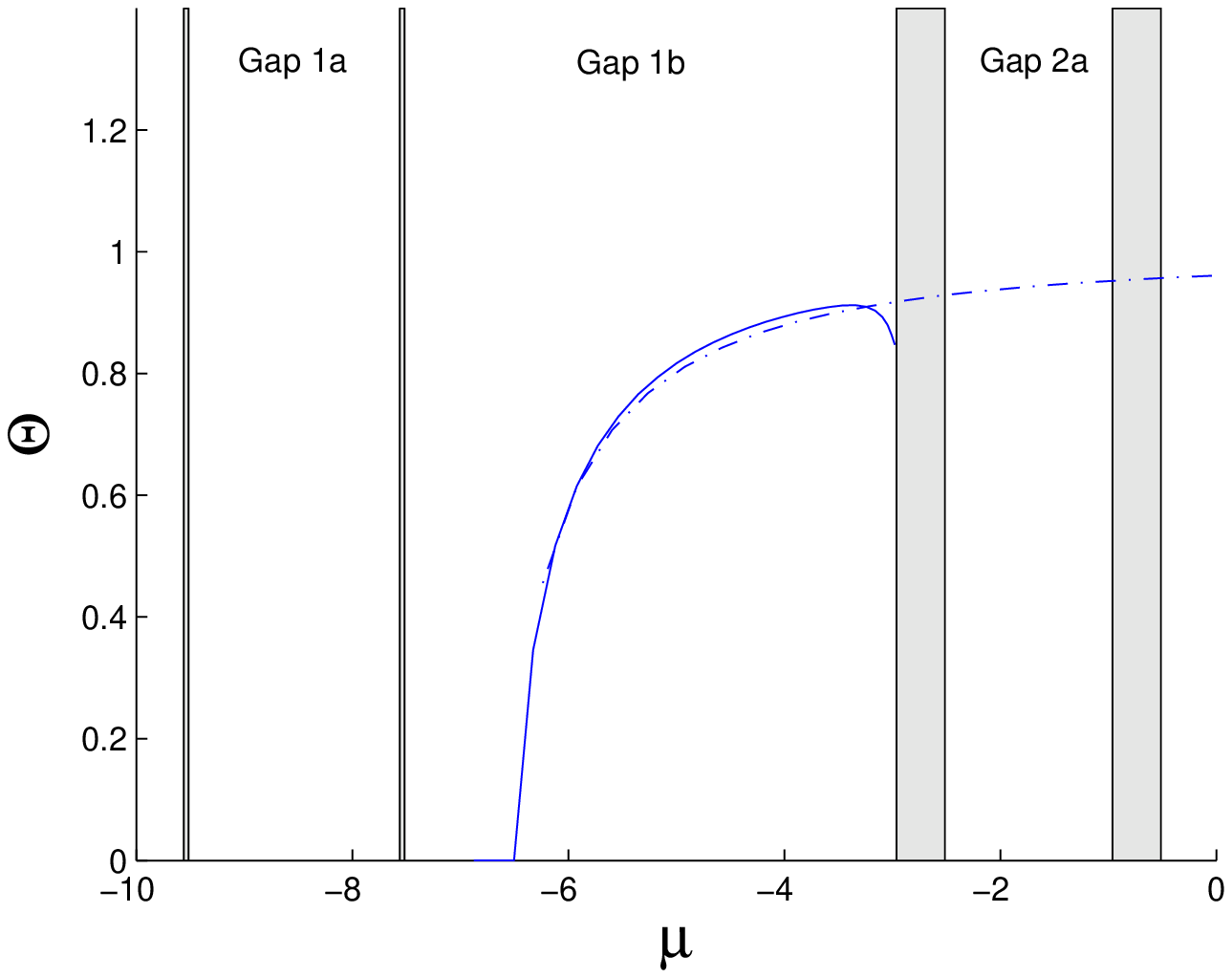}}
\caption{(Color online) The same as in Fig. \protect\ref{AA_solitons_family}%
, but for families of solitons in the symmetric repulsion-repulsion model ($%
\protect\lambda _{1}=\protect\lambda _{2}=-1$, $\Delta =0$). The portion of
the asymmetric-soliton branch in subgap 2a is an artifact of the variational
approximation.}
\label{RR_solitons_family}
\end{figure}

\begin{figure}[tbp]
\includegraphics[width=3in]{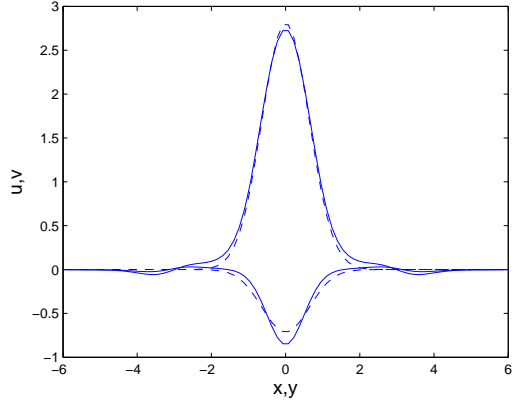}
\caption{(Color online) Comparison of profiles of asymmetric solitons, as
predicted by the VA and found in the numerical form (dashed and solid lines,
respectively), in the symmetric repulsive-repulsive model. In this case, $%
\protect\mu =-4$, and $N_{u}\approx 10.5,N_{v}\approx 0.6.$}
\label{va_vs_num_RR}
\end{figure}

\begin{figure}[tbp]
\includegraphics[width=3in]{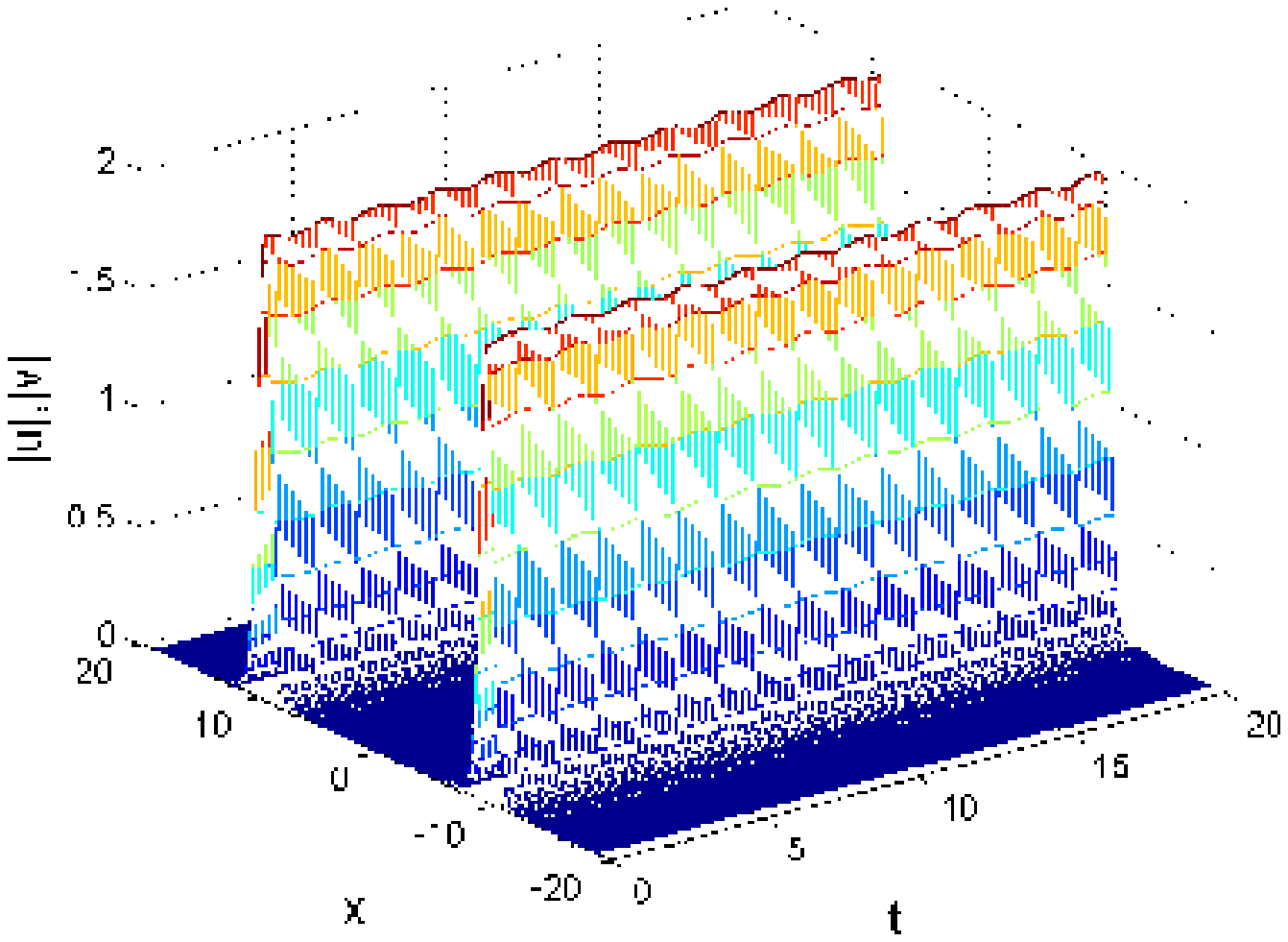}
\caption{(Color online) Evolution of a perturbed symmetric embedded soliton
in the symmetric repulsion-repulsion model is shown in the cross-section
along $x=0$. The chemical potential of the unperturbed soliton is $\protect%
\mu =-7.54$, placing it into the Bloch zone separating subgaps 1a and 1b.
The norm of the soliton is $N\approx 8$. The soliton is obviously stable.}
\label{embedded_evol_RR}
\end{figure}

A generic example of families of solitons found in the RR model is shown in
Fig. \ref{RR_solitons_family}. In subgap 1b, the VA predicts the solutions
very accurately. For this case, the comparison of numerical and variational
asymmetric soliton profiles is displayed in Fig. \ref{va_vs_num_RR}. On the
other hand, the VA also predicts asymmetric solutions in subgap 2a, where
such solutions could not be found in the numerical form, therefore the
extension of the asymmetric branch into the latter subgap is an artifact of
the variational method. It is noteworthy too that the stable branch of
symmetric solitons crosses the Bloch band separating subgaps 1a and 1b, thus
providing for the first example of 2D embedded solitons in a self-defocusing
medium. These embedded solitons are stable, as illustrated by an example of
the evolution of a perturbed soliton shown in Fig. \ref{embedded_evol_RR}.

The situation in the RR system is closer to what was found for its 1D
counterpart in Ref. \cite{we} (because collapse does not occur with
repulsive nonlinearity): the antisymmetric branch undergoes a supercritical
bifurcation, which gives rise to stable asymmetric solitons, while the
symmetric branch does not bifurcate and remains always stable (the VK
criterion does not apply to gap solitons in models with the repulsive
nonlinearity). Note that, like in Fig. \ref{AA_solitons_family}, the
bifurcation point is very accurately predicted by the VA. The asymmetric
solitons are stable whenever they exist, hence the bistability between
symmetric and antisymmetric or asymmetric solitons takes place here. The
antisymmetric solitons, destabilized by the SSB bifurcation, transform
themselves into persistent localized breathers, see an example in Fig. \ref%
{antisymm_RR_breather}. The breather features oscillations in its two
components, with equal amplitudes and phase shift $\pi /2$ between them.

\begin{figure}[tbp]
\centering\includegraphics[width=3in]{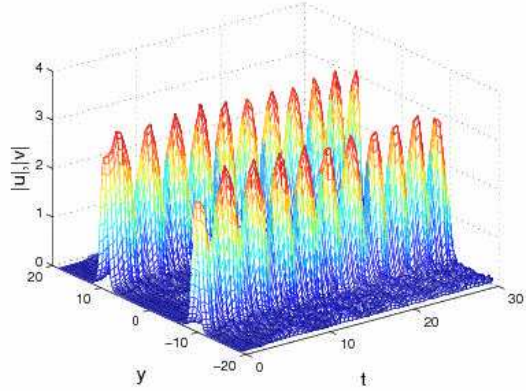}
\caption{(Color online) Spontaneous transformation of an unstable
antisymmetric soliton (with $\protect\mu =-3.5$) into a persistent breather,
in the symmetric repulsion-repulsion model ($\protect\lambda _{1}=\protect%
\lambda _{2}=-1$, $\Delta =0$).}
\label{antisymm_RR_breather}
\end{figure}

\begin{figure}[tbp]
\centering\includegraphics[width=3in]{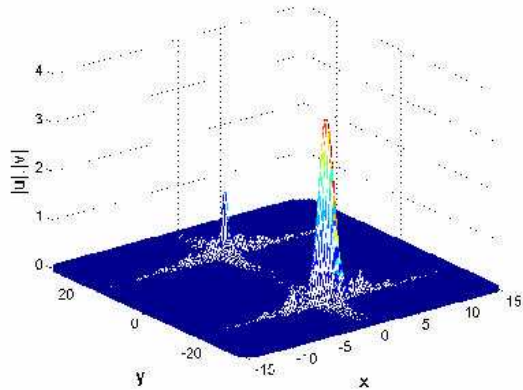}
\caption{(Color online) An example of a stable asymmetric soliton found, in
the symmetric repulsion-repulsion system, in subgap 2a, with $\protect\mu =1$%
.}
\label{sol_RR_2nd_gap_example}
\end{figure}

Lastly, stable asymmetric solitons have also been found in higher bandgaps,
an example of which is given in Fig. \ref{sol_RR_2nd_gap_example}.
Typically, the solitons in higher bandgaps are less tightly bound, featuring
well-pronounced sidelobes.

\section{Numerical results: vortices}

\subsection{Symmetric attraction-attraction system}

\begin{figure}[tbp]
\subfigure[]{\includegraphics[width=3in]{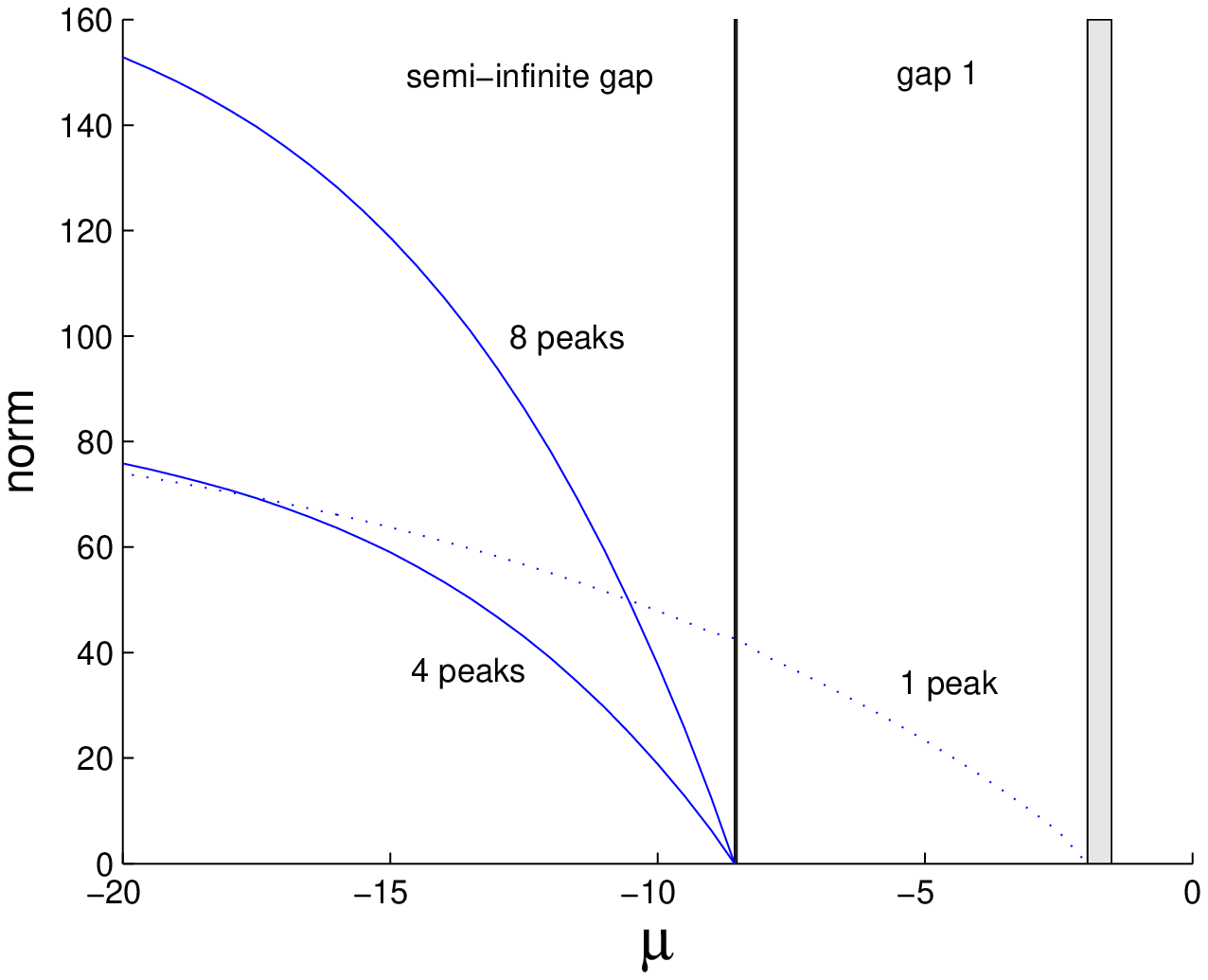}} \subfigure[]{%
\includegraphics[width=3in]{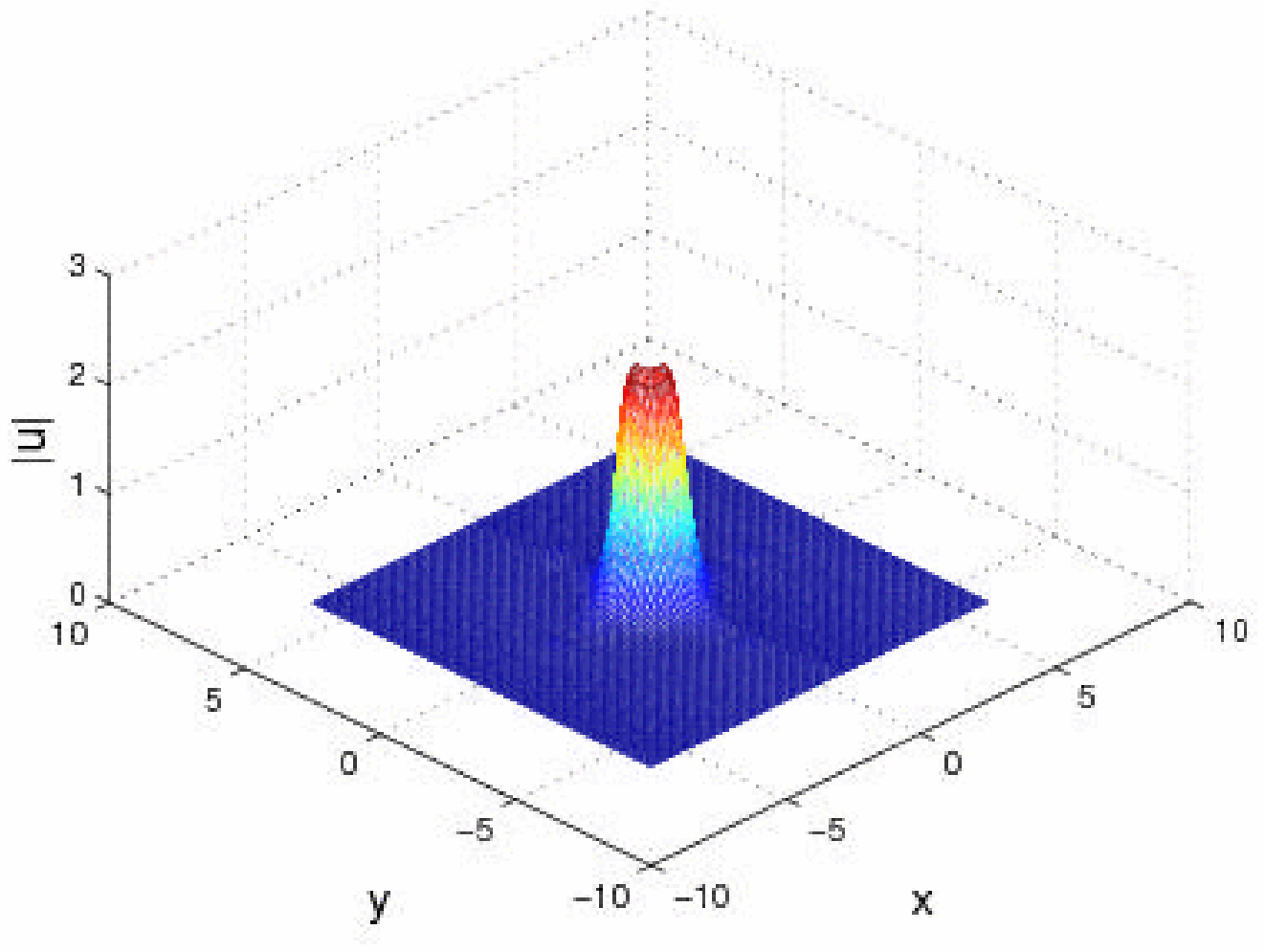}} \subfigure[]{%
\includegraphics[width=3in]{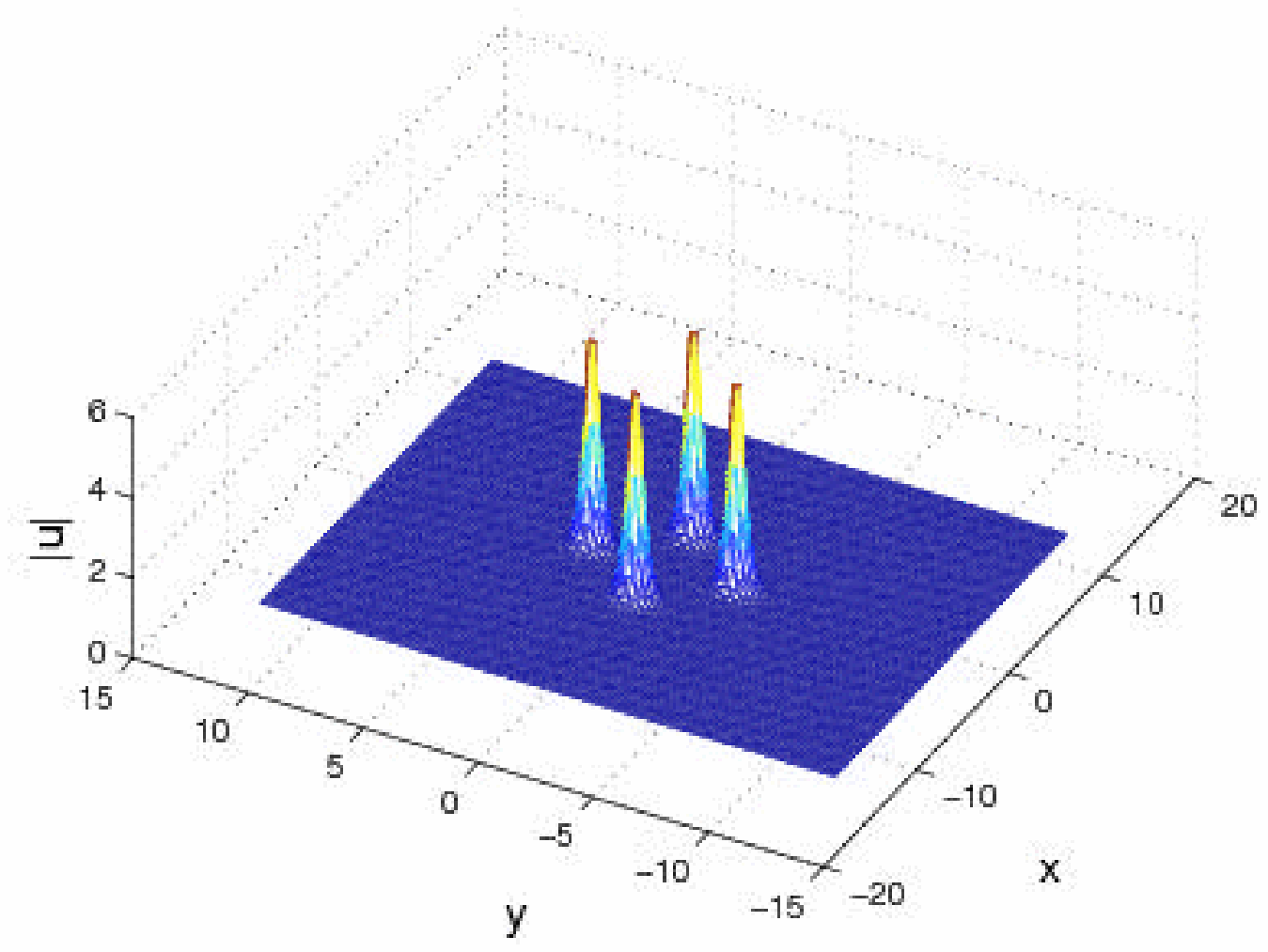}} \subfigure[]{%
\includegraphics[width=3in]{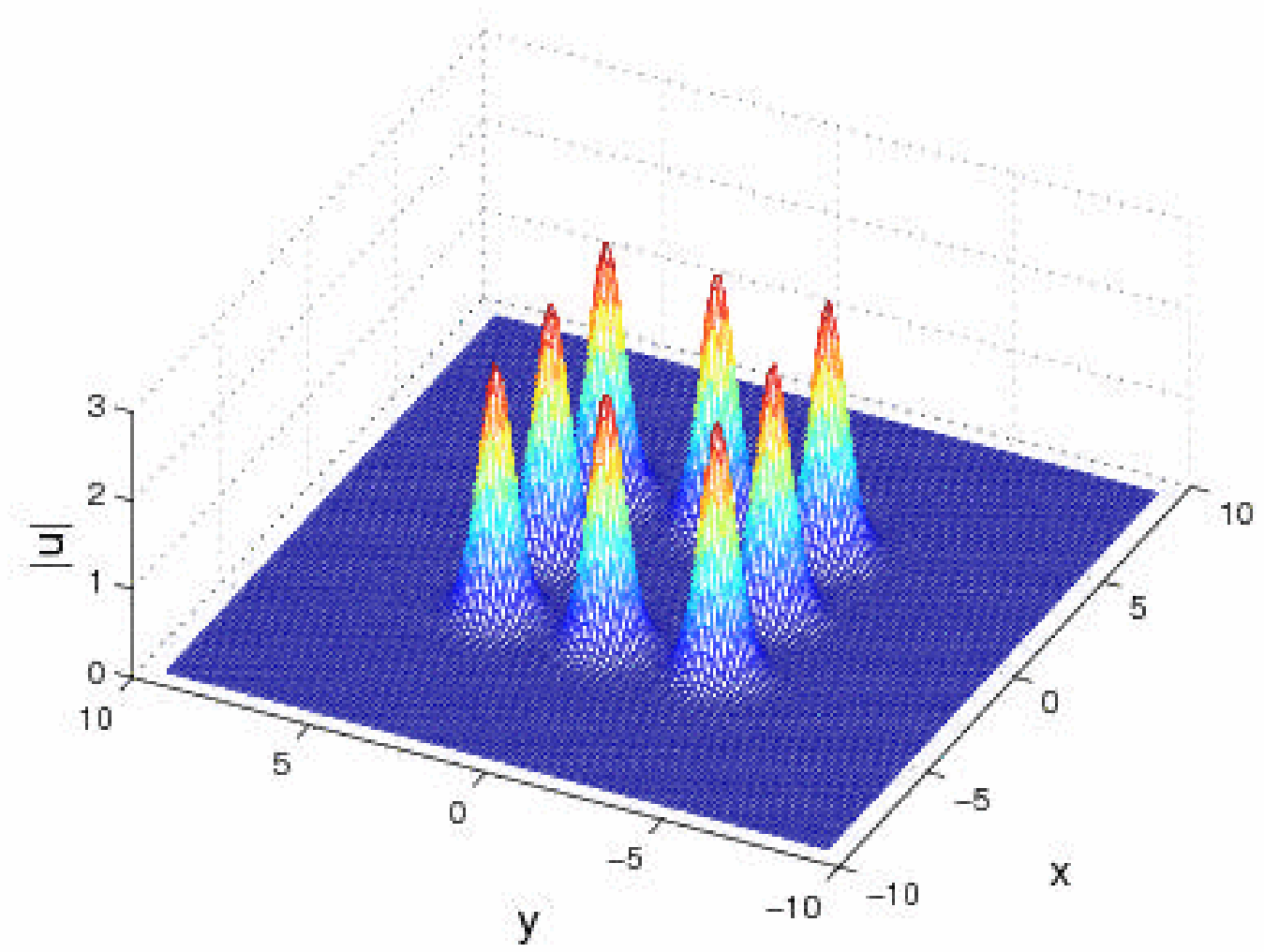}}
\caption{(Color online) Localized vortices with topological charge $S=1$ in
the single-component model with attraction ($\protect\lambda =1$). (a) Three
solution families (unstable single-peak crater-shaped vortices, and stable
4- and 8-peak vortex complexes) are represented by the dependence of their
norm on the chemical potential. (b)-(d) Examples of the three species of
localized vortices, with $\left( \protect\mu =-4.5,~N\approx 20\right) $, $%
\left( \protect\mu =-20,~N\approx 76\right) $, and $\left( \protect\mu %
=-12,~N\approx 78\right) $, respectively.}
\label{single_vx_A}
\end{figure}

In the model based on the single-component GPE in two dimensions
with the OL and attractive nonlinearity, stable solutions in the
form of localized vortices with topological charge (``spin") $S=1$
were reported in Refs. \cite{Baizakov} and \cite{Jianke}, and
their (also stable) counterparts with $S>2$ were found in Ref.
\cite{Sakaguchi2}. Before proceeding to the new problem of the SSB
in two-component vortices, it is relevant to recapitulate basic
results obtained for vortex solitons in the single-component
model. Figure \ref{single_vx_A} shows a generic example of
families of vortex-soliton solutions with $S=1$ in this model. The
most compact ``crater-shaped" vortices, which feature a single
peak with a hole in the center, are always unstable. However, the
vortices built as 4- and 8-peak complexes, with phase shifts,
respectively, $\pi /2$ or $\pi /4$ between the peaks (which
corresponds to the net phase circulation of $2\pi $, i.e., $S=1$),
constitute entirely stable families (Fig. \ref{single_vx_A}
includes examples of all the three species of localized vortices).
These complexes are quite similar to examples of stable vortices
reported in Refs. \cite{Baizakov} and \cite{Jianke}.

\begin{figure}[tbp]
\subfigure[]{\includegraphics[width=3in]{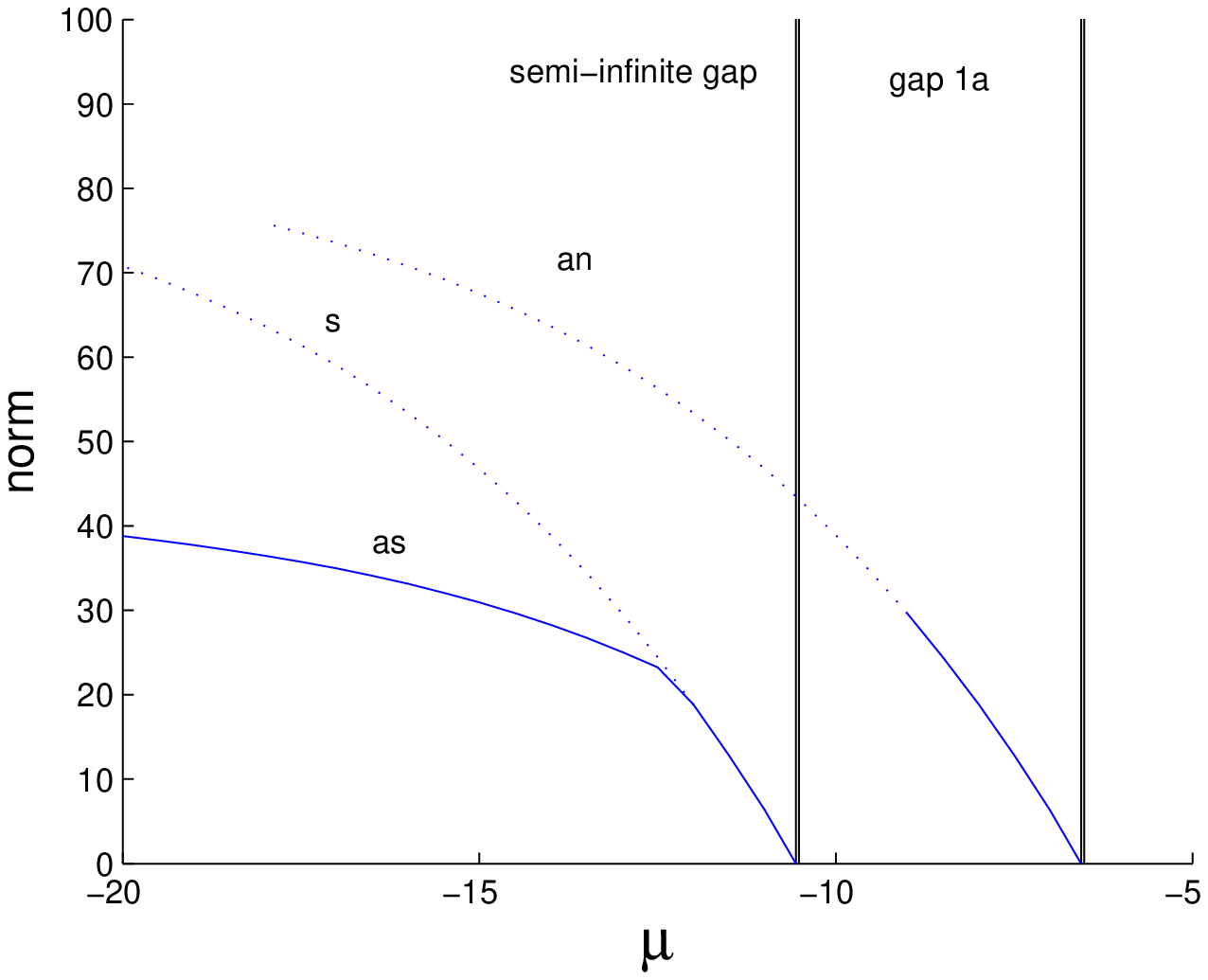}} \subfigure[]{%
\includegraphics[width=3in]{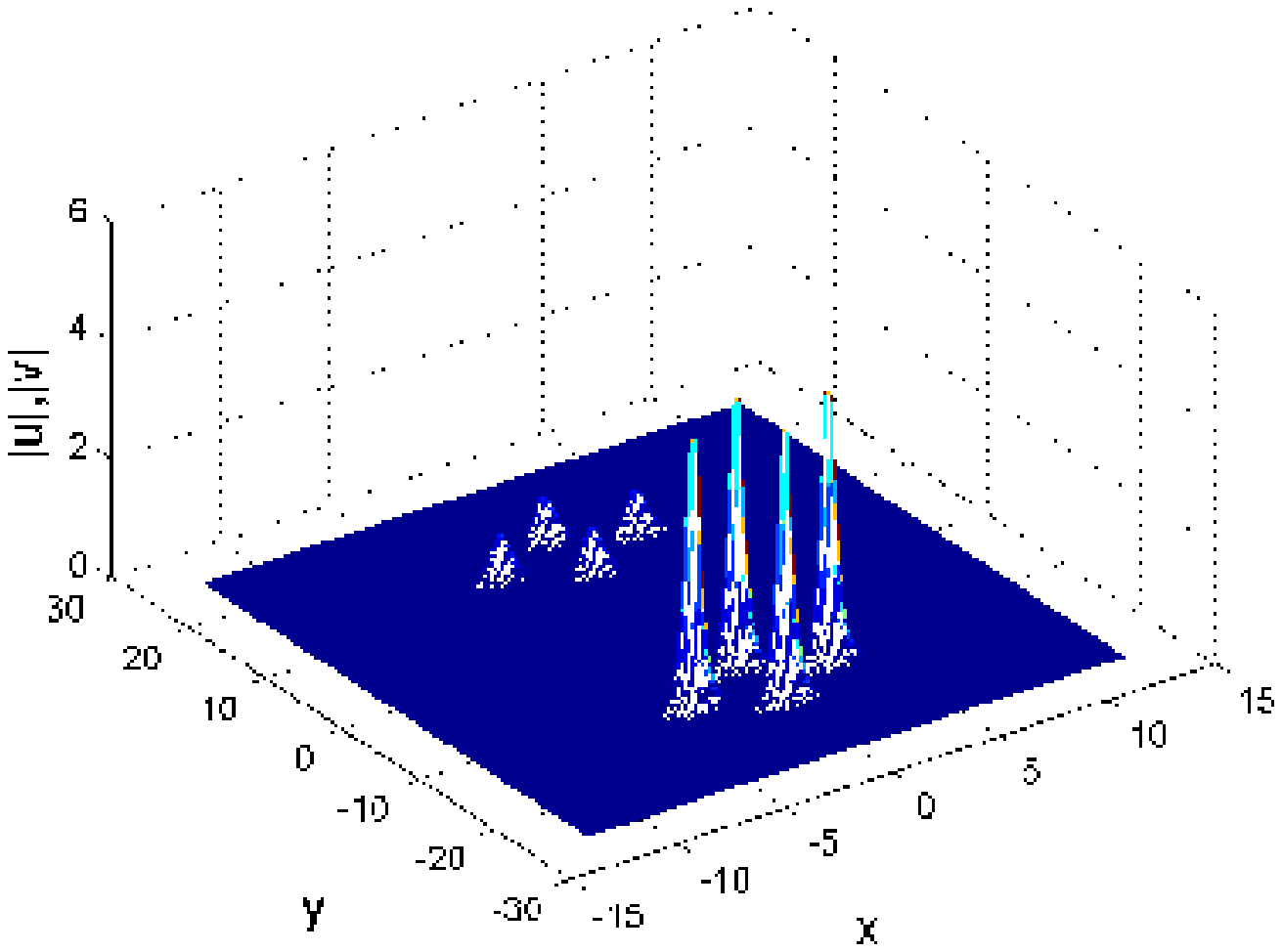}}
\caption{(Color online) 4-peak vortices with topological charge $S=1$, in
the symmetric attraction-attraction system, with $\protect\kappa =2$. (a)
Families of the symmetric, antisymmetric, and asymmetric vortex solutions
represented by $N(\protect\mu )$ curves. (b) An example of a stable
asymmetric vortex, for $\protect\mu =-18$ and $N=36.5$.}
\label{vx_AA_4hump}
\end{figure}

We have found the SSB of two-component vortex solitons in the symmetric AA
system ($\lambda _{1}=\lambda _{2}=+1$, $\Delta =0$). Families of 4- and
8-peak vortices of the symmetric, antisymmetric and asymmetric types, found
in the coupled system, are shown in Figs. \ref{vx_AA_4hump} and \ref%
{vx_AA_8hump}, along with examples of respective stable asymmetric vortices.

\begin{figure}[tbp]
\subfigure[]{\includegraphics[width=3in]{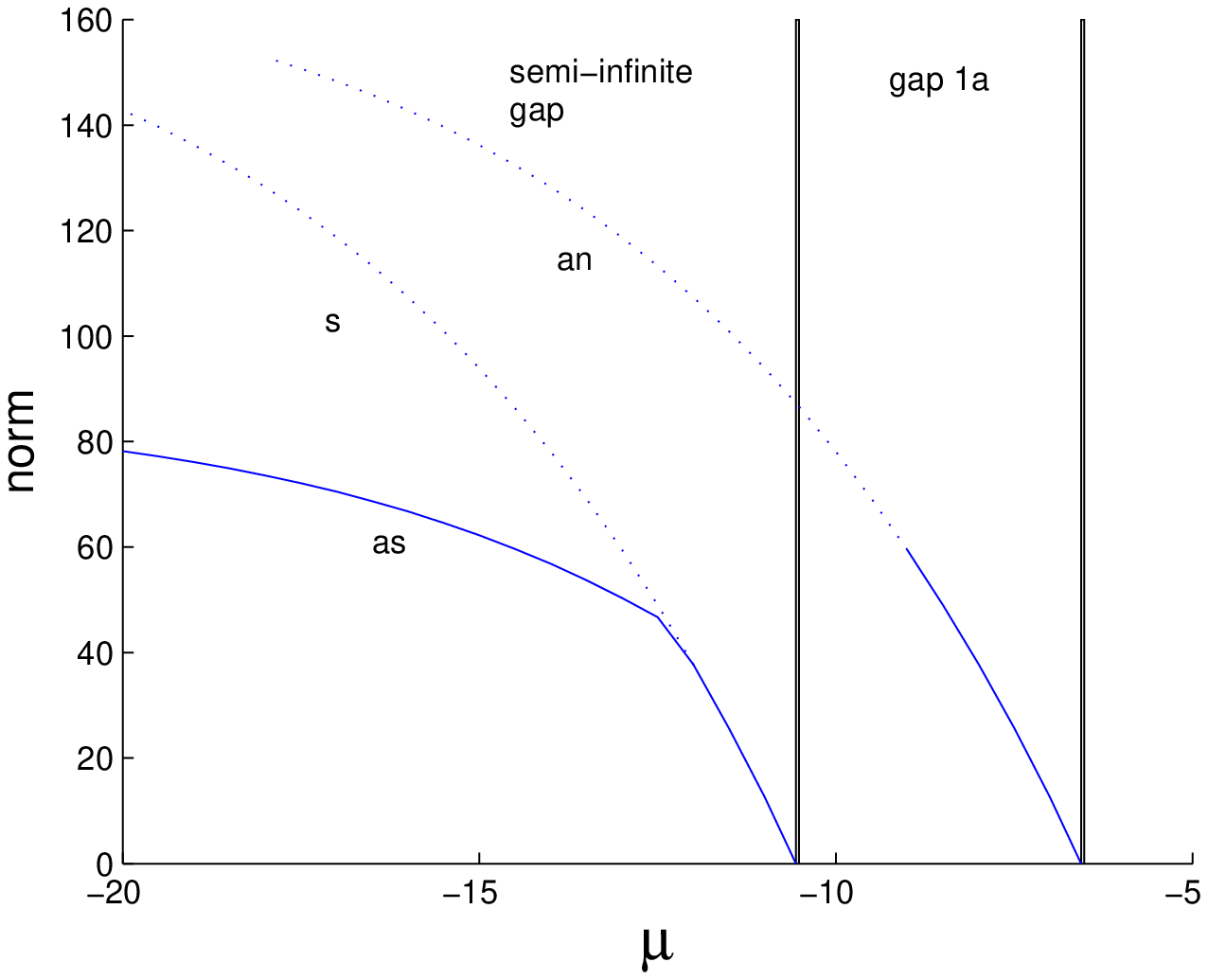}} \subfigure[]{%
\includegraphics[width=3in]{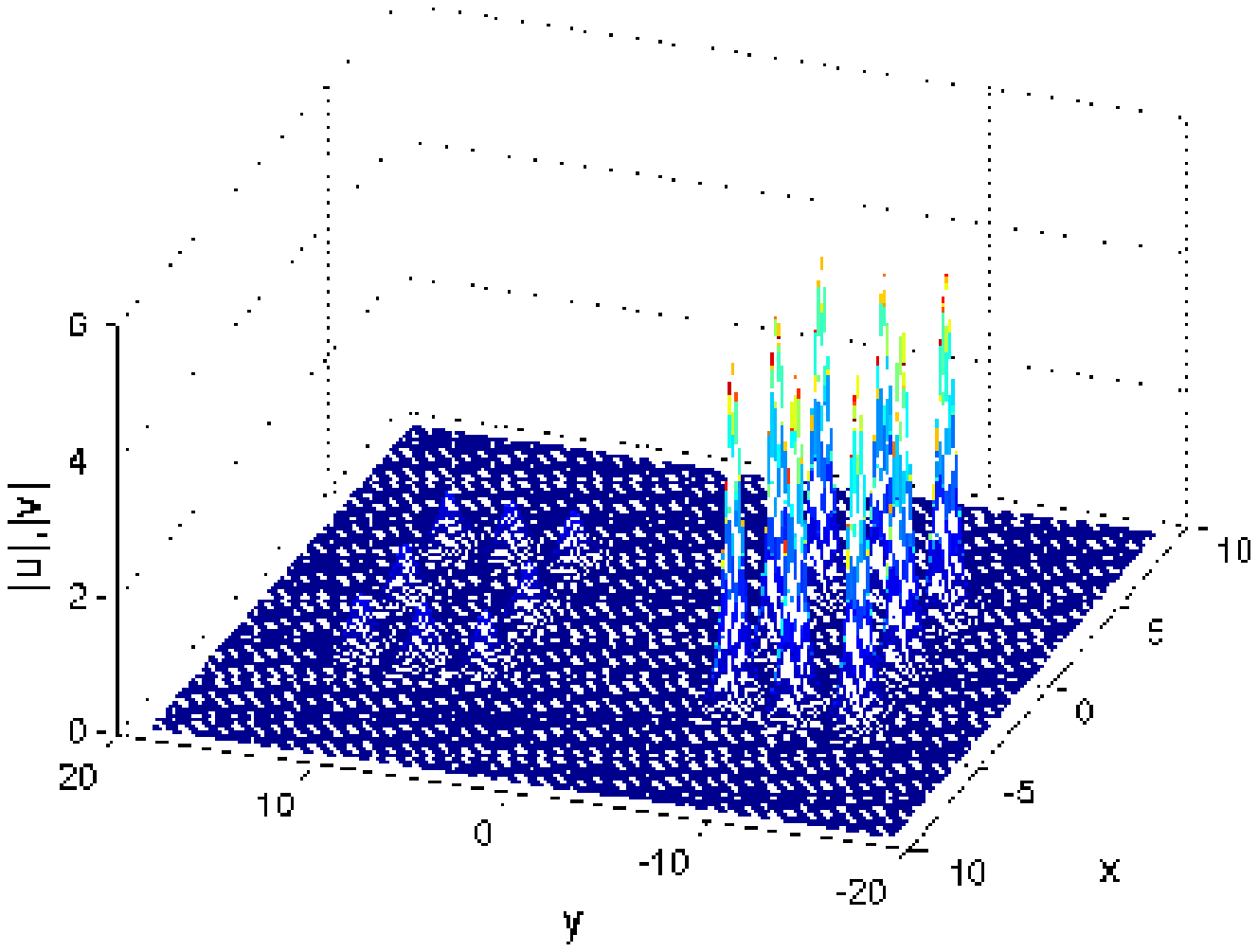}}
\caption{(Color online) The same as in Fig. \protect\ref{vx_AA_4hump}, but
for 8-peak vortices. In panel (b), an example of the stable asymmetric
vortex is displayed for $\protect\mu =-18$, $N=73.5$.}
\label{vx_AA_8hump}
\end{figure}

The stability of these solutions was identified, as above, by direct
simulations. Symmetric vortices get destabilized by the bifurcation and tend
to spontaneously rearrange into asymmetric ones, which emerge as stable
solutions, while antisymmetric vortices do not bifurcate. At large values of
the norm, the localized vortices are subject to collapse.

\subsection{Symmetric repulsion-repulsion system}

\begin{figure}[tbp]
\subfigure[]{\includegraphics[width=3in]{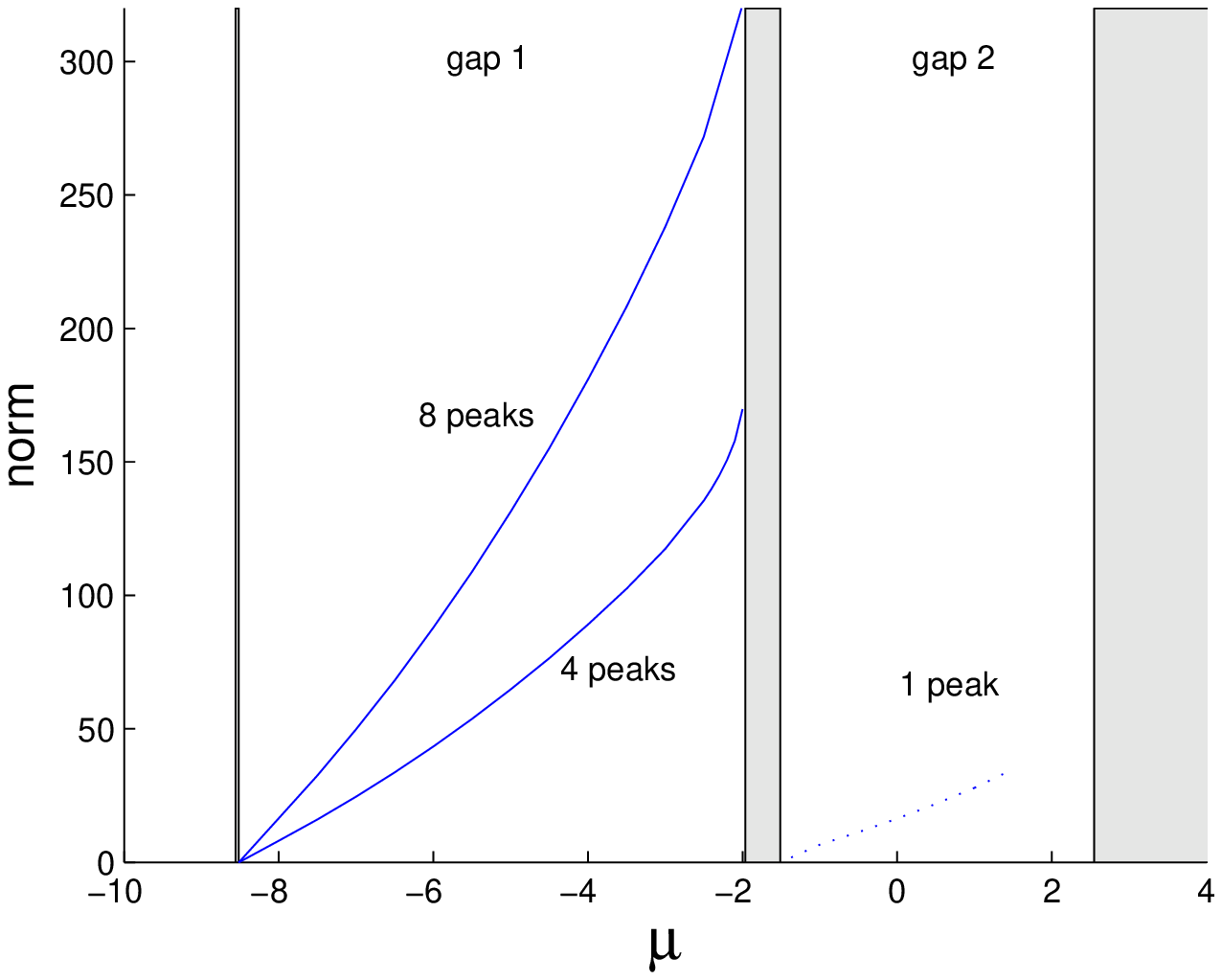}} \subfigure[]{%
\includegraphics[width=3in]{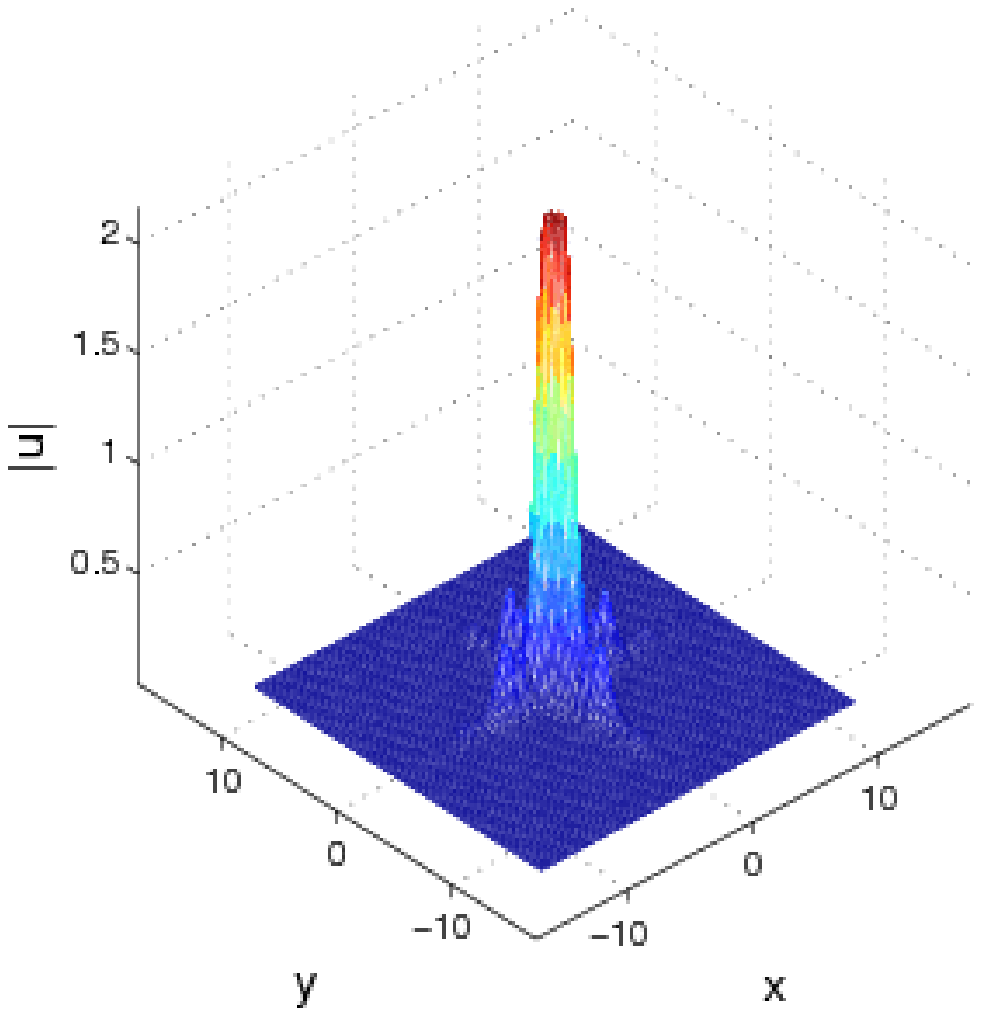}} \subfigure[]{%
\includegraphics[width=3in]{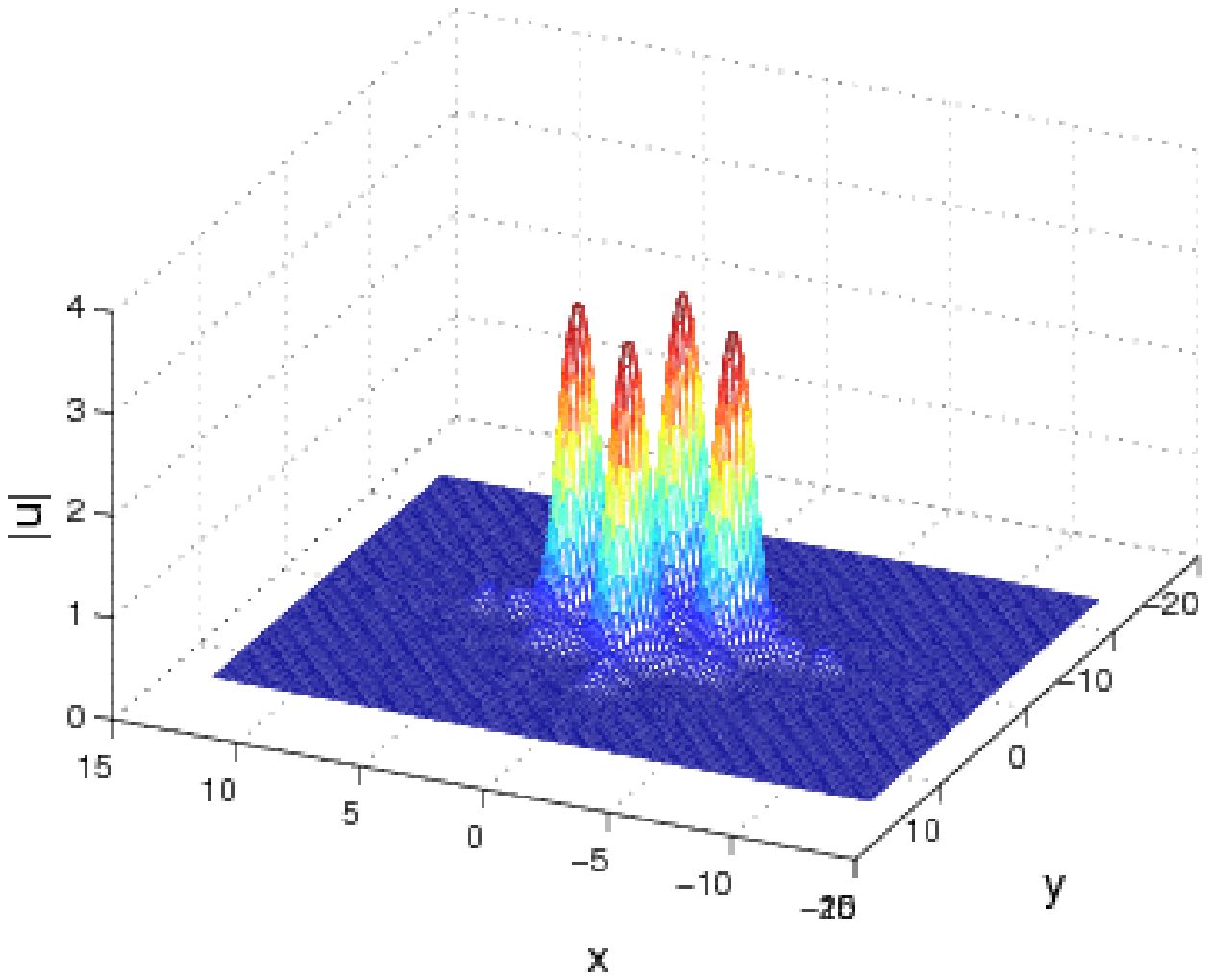}} \subfigure[]{%
\includegraphics[width=3in]{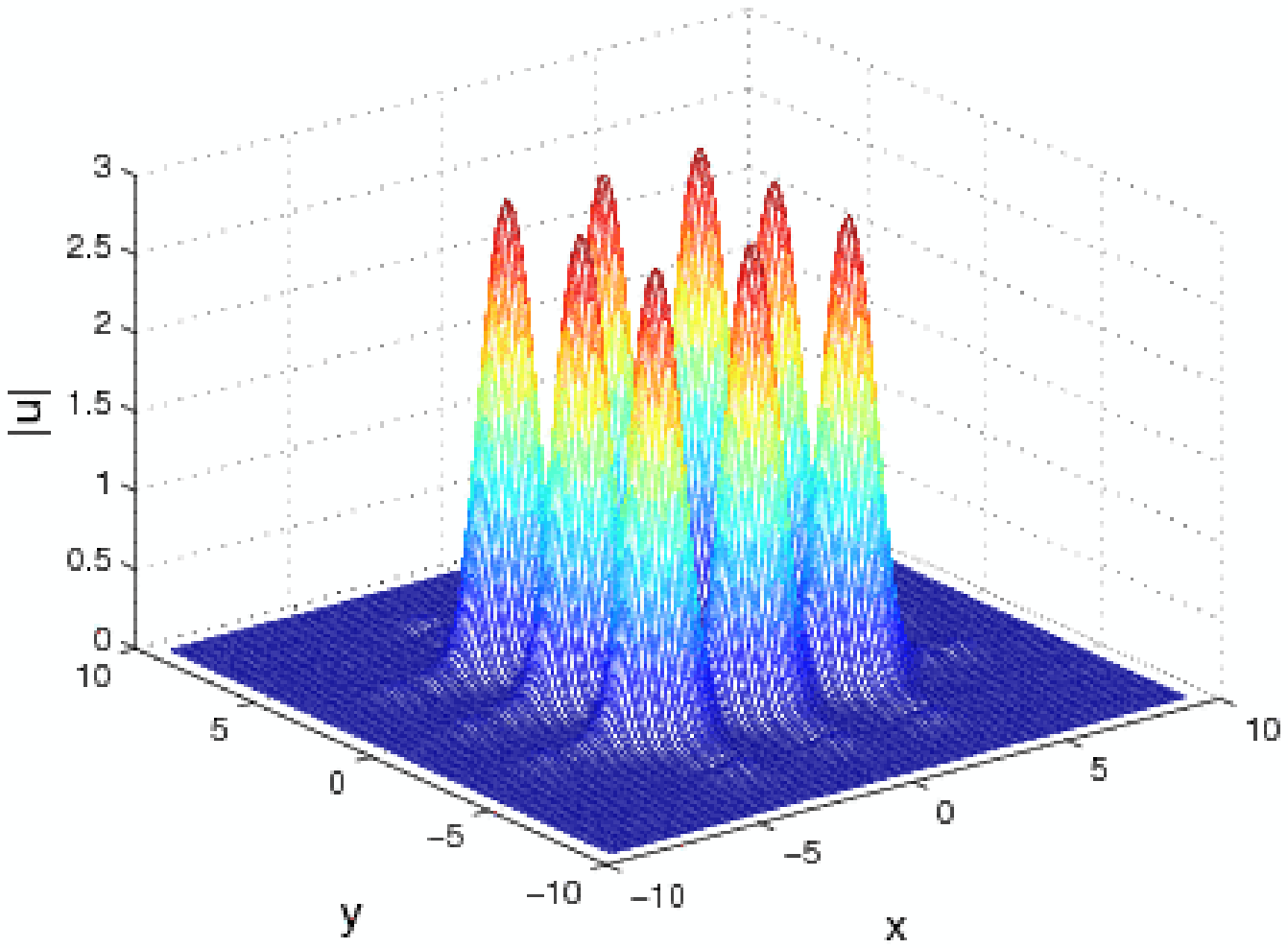}}
\caption{(Color online) Localized vortices in the single-component model
with repulsion ($\protect\lambda =-1$). (a) Norm versus the chemical
potential for three species of vortex solitons: single-peak, 4-peak, and
8-peak ones. (b)-(d): Examples of vortices of these types, with $\left(
\protect\mu =1.5,~N\approx 35\right) $, $\left( \protect\mu =-2.5,~N\approx
135\right) $, and $\left( \protect\mu =-3.5,~N\approx 208\right) $,
respectively.}
\label{single_vx_R}
\end{figure}

\begin{figure}[tbp]
\subfigure[]{\includegraphics[width=3in]{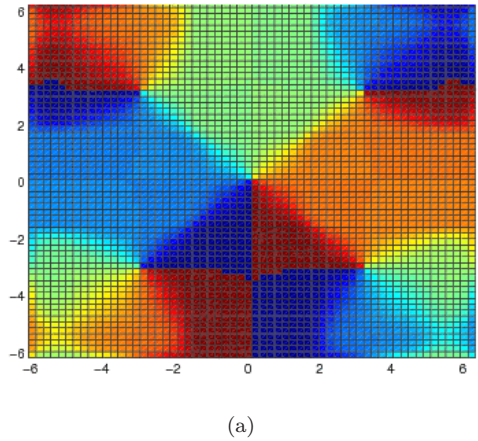}}
\caption{(Color online) The phase pattern in the 4-peak vortex complex in
the single-component repulsive model, whose density profile is displayed in
Fig. \protect\ref{single_vx_R}(c).}
\label{single_vx_R_4_phase}
\end{figure}

In the repulsive model with the OL, localized vortices may be treated as a
species of 2D solitons of the gap type \cite{Ostrovskaya2,Sakaguchi}. First,
in Fig. \ref{single_vx_R} we present families of vortex solitons in the
single-component model. Similar to what was shown above for the attractive
model, the single-peak (crater-shaped) vortices are unstable, while
multiple-peak vortical complexes are stable. However, in the repulsive model
only the (unstable) single-peak and (stable) 8-peak structures carry
topological charge $S=1$, while the 4-peak entity is a complex bound state
of several vortices. Indeed, the phase distribution in the latter state,
displayed in Fig. \ref{single_vx_R_4_phase}, suggests that it may be
interpreted as a structure built of of five vortices: one, with $S=1$, is
located in the center, being surrounded by four constituent vortices, each
carrying charge $S=-1$.

\begin{figure}[tbp]
\subfigure[]{\includegraphics[width=3in]{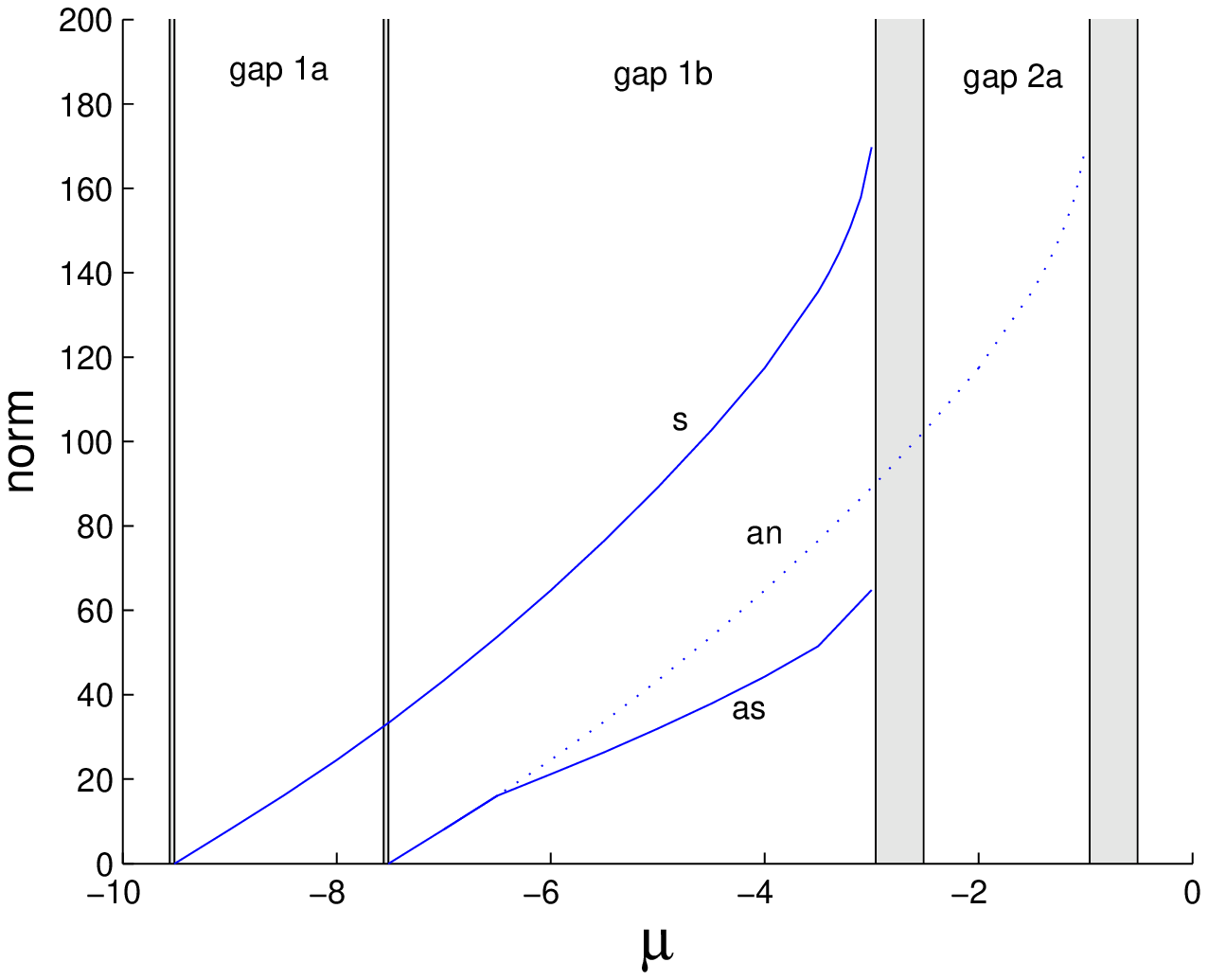}} \subfigure[]{%
\includegraphics[width=3in]{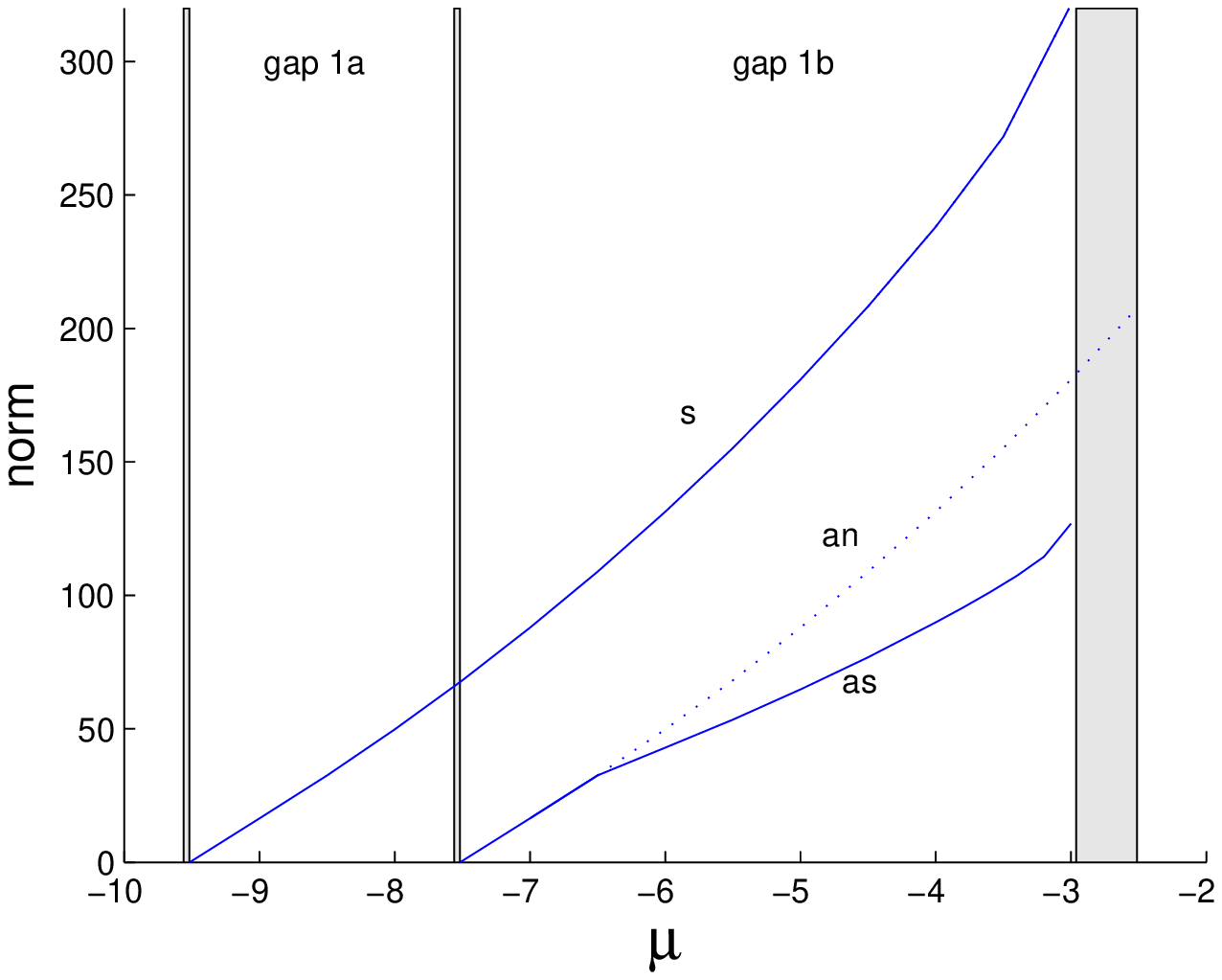}}
\caption{(Color online) Families of 4-peak (a) and 8-peak (b) vortices in
the symmetric repulsion-repulsion system.}
\label{vx_RR}
\end{figure}

In the coupled (two-component) RR system, we observed SSB in both types of
stable vortex states, 4-peak and 8-peak ones, as shown in Fig. \ref{vx_RR}.
Similar to what was demonstrated above for the solitons, the asymmetric
branch bifurcates from the antisymmetric one, while the family of the
symmetric vortices does not bifurcate and remains entirely stable, thus
giving rise to the bistability, together with the stable asymmetric
vortices. Antisymmetric two-component vortices destabilized by the SSB
bifurcation develop into persistent breathers. A noteworthy feature of Fig. %
\ref{vx_RR} is the fact that the branch of symmetric solutions (both 4- and
8-peak ones) crosses the Bloch band between subgaps 1a and 1b, which
provides for the first example of (stable) \textit{embedded vortex solitons}.

\begin{figure}[tbp]
\caption{(Color online) A stable asymmetric vortex with topological charge $%
S=2$, found in the symmetric repulsion-repulsion system. In this case, $%
\protect\mu =1$ (which falls in subgap 2a). Norms of the components are $%
N_{u}\approx 272$, $N_{v}\approx 13$. Note the difference in the $x$- and $y$%
- scales in the figure (in fact, the vortex is invariant with respect to $%
x\leftrightarrow y$).}\includegraphics[width=3in]{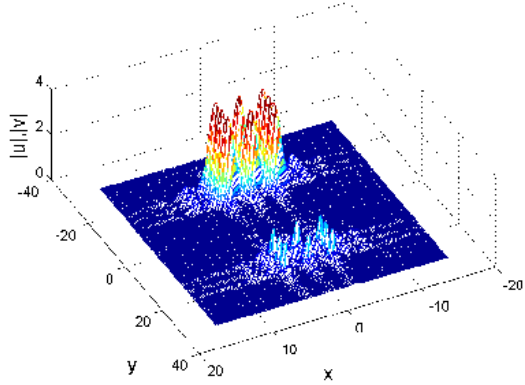}
\label{vx_RR_n2_example}
\end{figure}

Stable asymmetric vortices residing in higher bandgaps, as well as stable
vortices with higher vorticity, $S>1$, have been found too. In particular,
Fig. \ref{vx_RR_n2_example} displays an example of a stable 8-peaked vortex
with $S=2$, found in subgap 2a.

\section{Asymmetric models}

As explained in Introduction, the symmetry of coupled equations (\ref%
{the_model_2d}) may be broken by opposite signs of the nonlinearity in the
two cores (in the AR system, with $\lambda _{1}=-\lambda _{2}=+1$), or by
mismatches between the lattices in them, $\Delta _{1,2}\neq 0$. We did not
aim to perform an exhaustive analysis of the asymmetric models, but examples
of stable solitons in these models have been found.

\begin{figure}[tbp]
\subfigure[]{\includegraphics[width=3in]{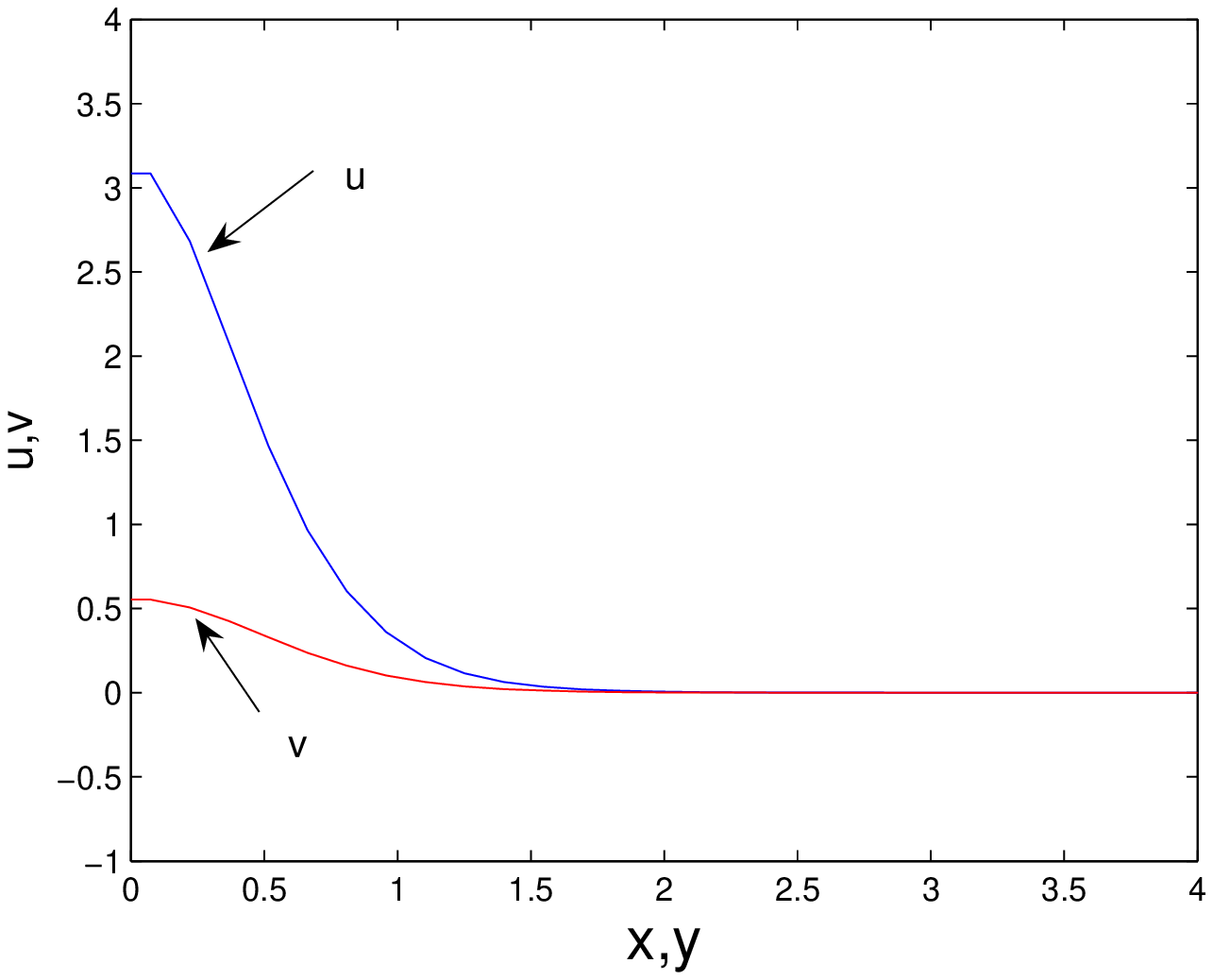}} \subfigure[]{%
\includegraphics[width=3in]{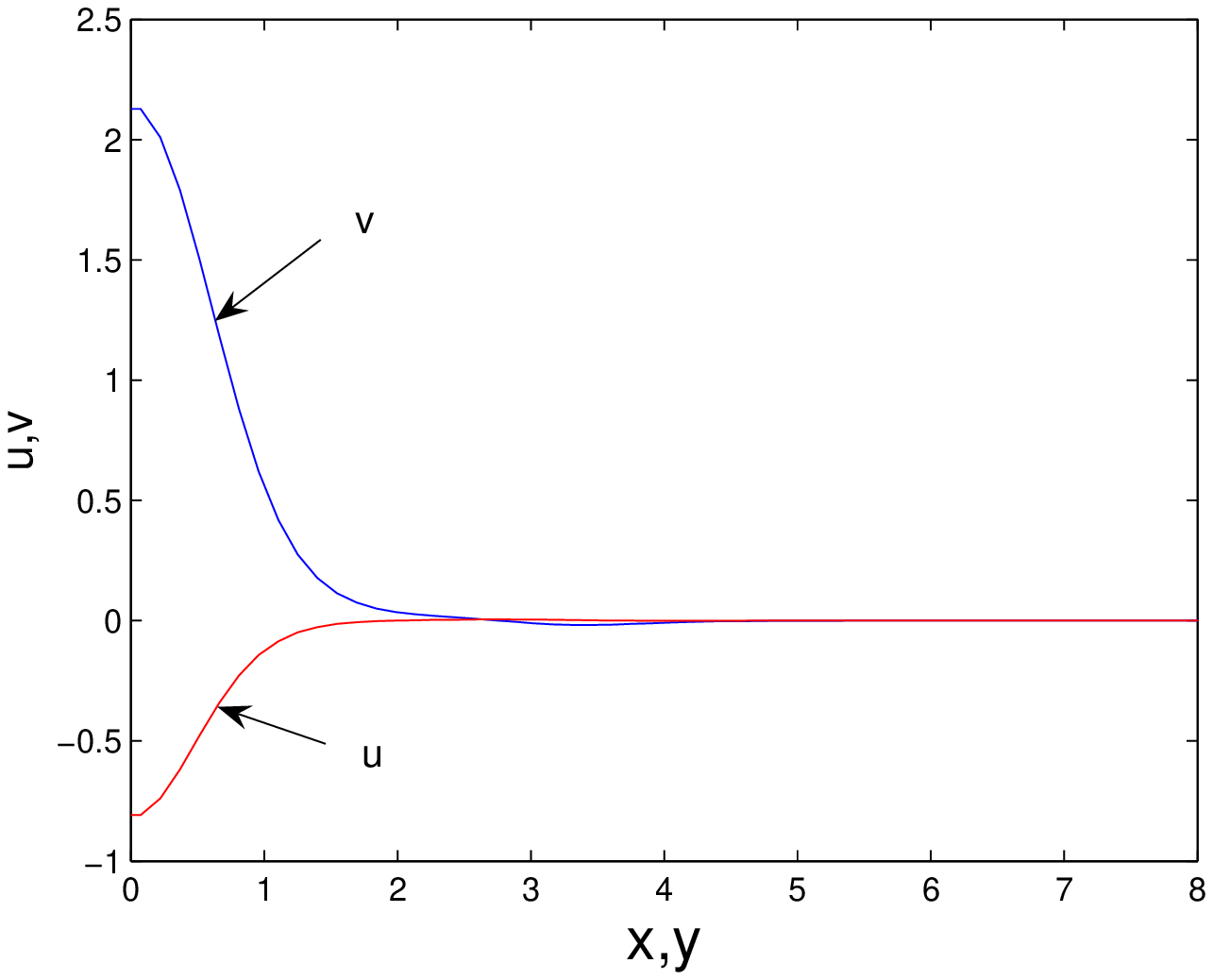}}
\caption{(Color online) Stable solitons in the coupled attraction-repulsion
system ($\protect\lambda _{1}=-\protect\lambda _{2}=1,~\Delta _{1,2}=0$).
The solitons' profiles are even functions of $x$ and $y$, and are invariant
with respect to transformation $x\leftrightarrow y$. (a) A soliton with the
dominant self-attractive component. It has $\protect\mu \approx -13$ (which
falls in the semi-infinite gap), and $N_{u}\approx 5.7,N_{v}\approx 0.26$.
(b) A soliton with the dominant self-repulsive component, found with $%
\protect\mu \approx -5.7$ (which belongs to subgap 1a) and $N_{u}\approx
0.55,N_{v}\approx 5.35$.}
\label{AR_examples}
\end{figure}

Figure \ref{AR_examples} shows stable solitons in the AR system. Solutions
of two types have been found in it: one with a dominant self-repulsive
component sitting in subgap 1a, and another one with the dominant
self-attractive component found, as should be expected, in the semi-infinite
gap. Stability of the solitons have been verified by direct simulations.

Stable 8-peak vortices with $S=1$ have also been found in the AR model.
Vortices with dominant self-attractive and self-repulsive components are
located in the semi-infinite gap and subgap 1a, respectively. Their examples
are not shown here, as they are very similar to the 8-peak asymmetric
solitons found in the AA and RR models, that were reported above.

\begin{figure}[tbp]
\subfigure[]{\includegraphics[width=3in]{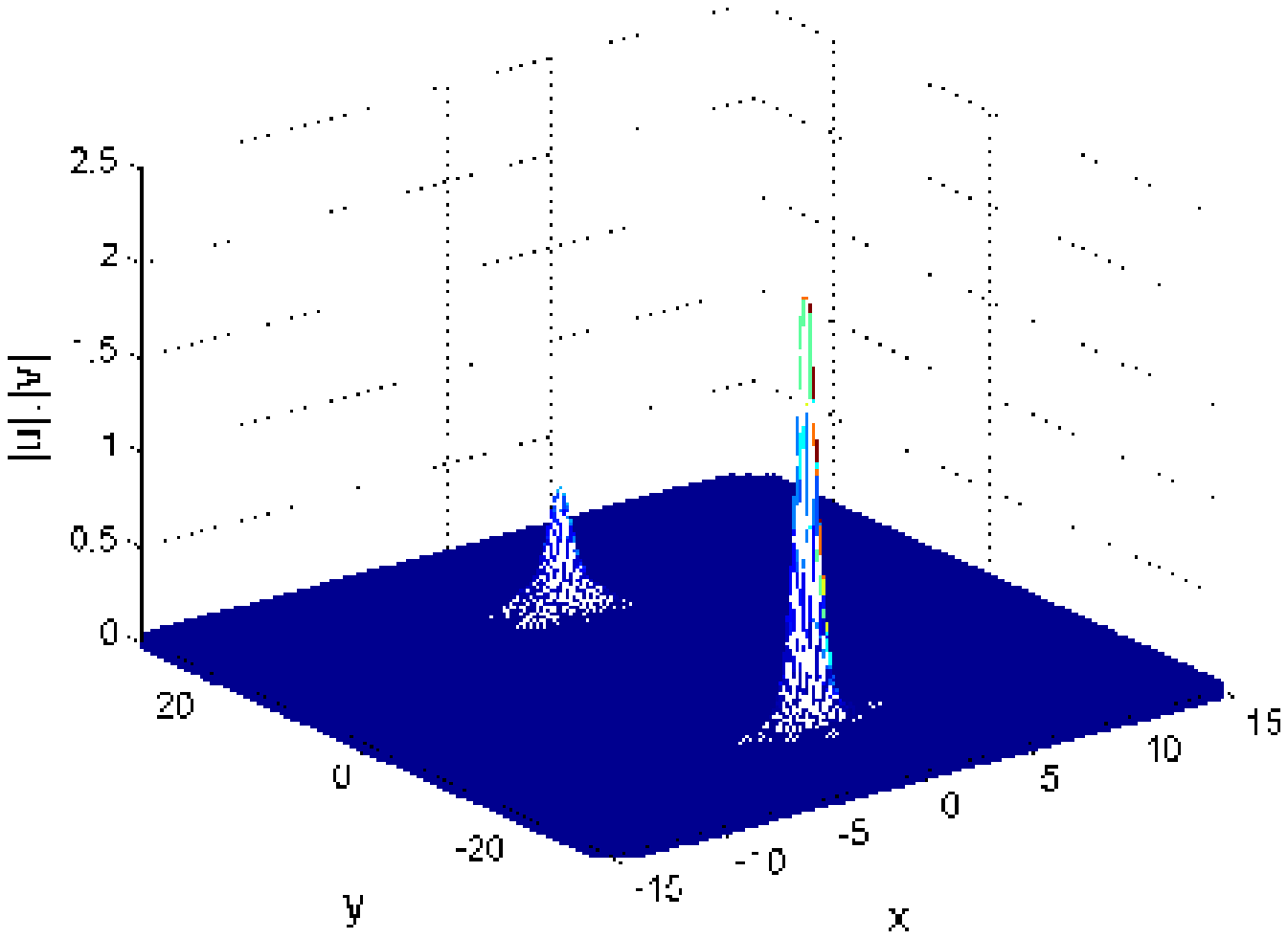}} \subfigure[]{%
\includegraphics[width=3in]{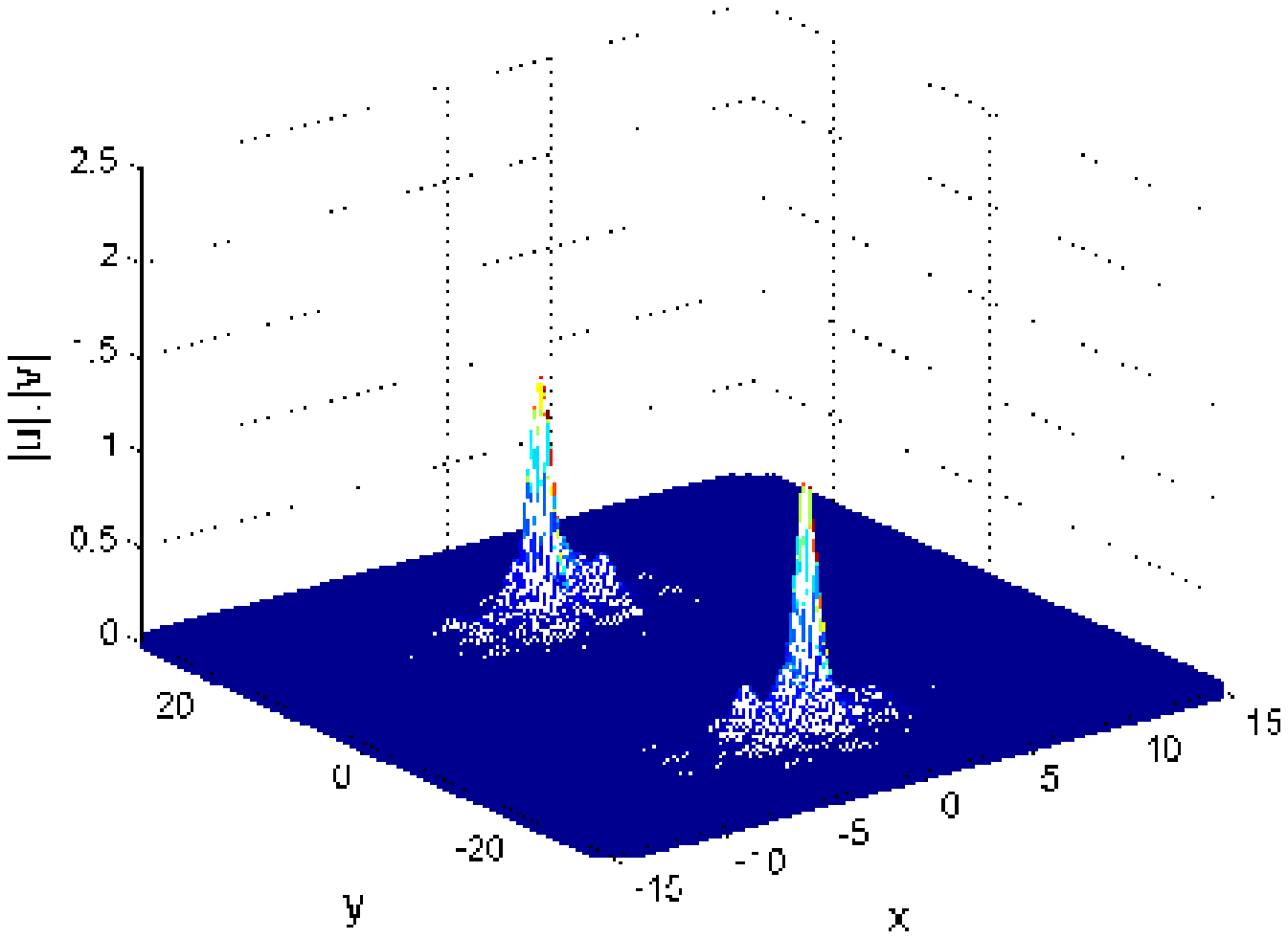}}
\caption{(Color online) Soliton profiles in the attraction-attraction model
with the maximum diagonal mismatch between the lattices in the cores, $%
\Delta _{1}=\Delta _{2}=\protect\pi $, for $\protect\kappa =10$, and $N=5$.
(a) A stable asymmetric soliton; (b) an unstable quasi-symmetric soliton.}
\label{offset_profiles_AA}
\end{figure}

We have also studied the system with the mismatch between the OLs,
concentrating on two most interesting cases, with $\Delta _{1}=\pi ,\Delta
_{2}=0$ or $\Delta _{1}=\Delta _{2}=\pi $ in Eqs. (\ref{the_model_2d}). In
either case, the phase mismatch was given the largest possible value ($\pi $%
). As in the 1D model with nonzero mismatch \cite{we}, quasi-symmetric and
asymmetric solutions have been identified, as states with $N_{u}=N_{v}$ and $%
N_{u}\neq N_{v}$, respectively. In the mismatched system, asymmetric
solitons have peaks in the two components located at the same position,
while in quasi-symmetric solutions the peaks are slightly separated. Typical
profiles of quasi-symmetric and asymmetric solutions in the misaligned AA
system are shown in Fig. \ref{offset_profiles_AA} for $\Delta _{1}=\Delta
_{2}=\pi $ (this choice corresponds to the largest diagonal mismatch).
Direct simulations demonstrate that the quasi-symmetric solitons are always
unstable in this case (an explanation of this feature is given below), while
the asymmetric solitons are stable, beneath the collapse threshold.

\begin{figure}[tbp]
\subfigure[]{\includegraphics[width=3in]{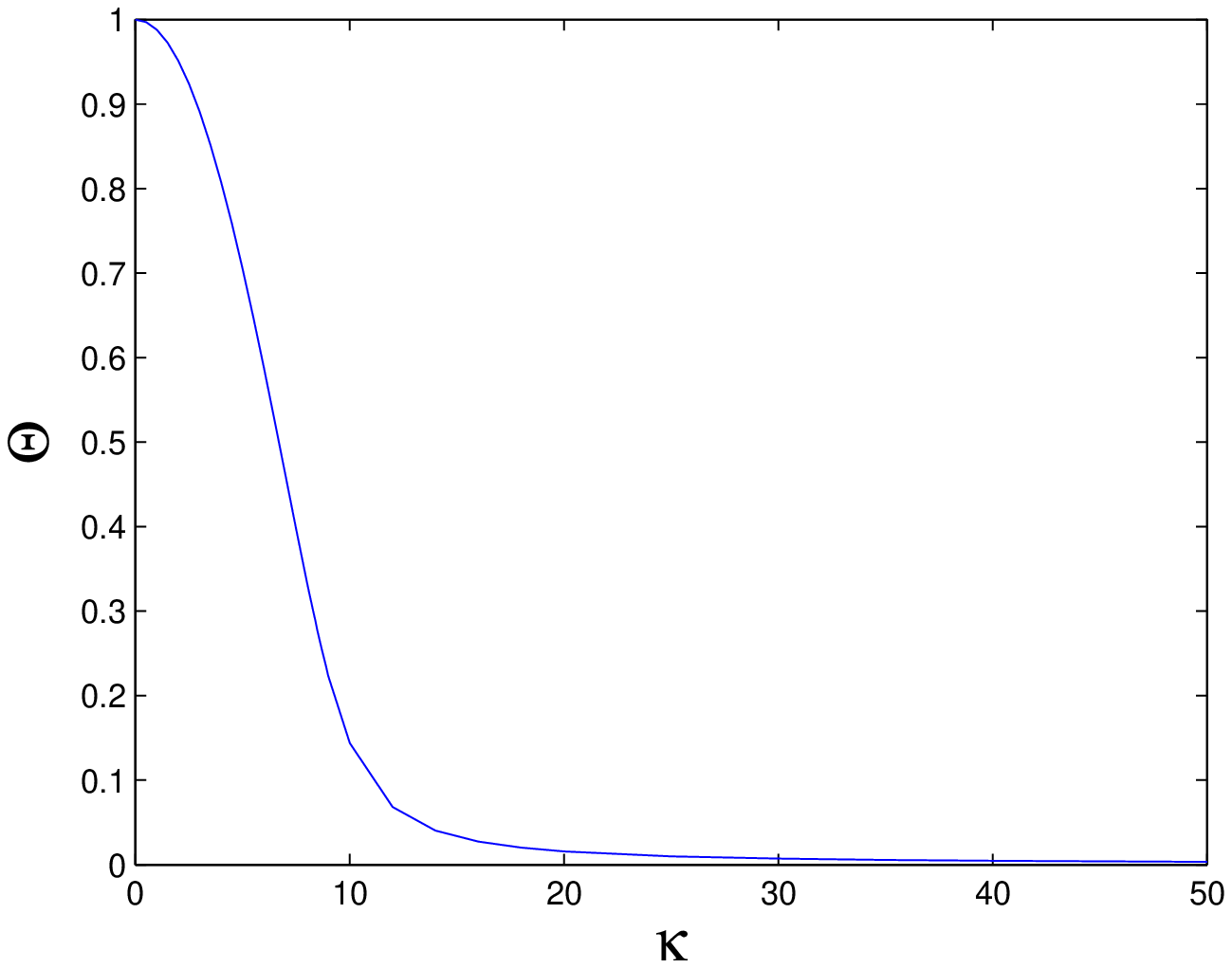}} \subfigure[]{%
\includegraphics[width=3in]{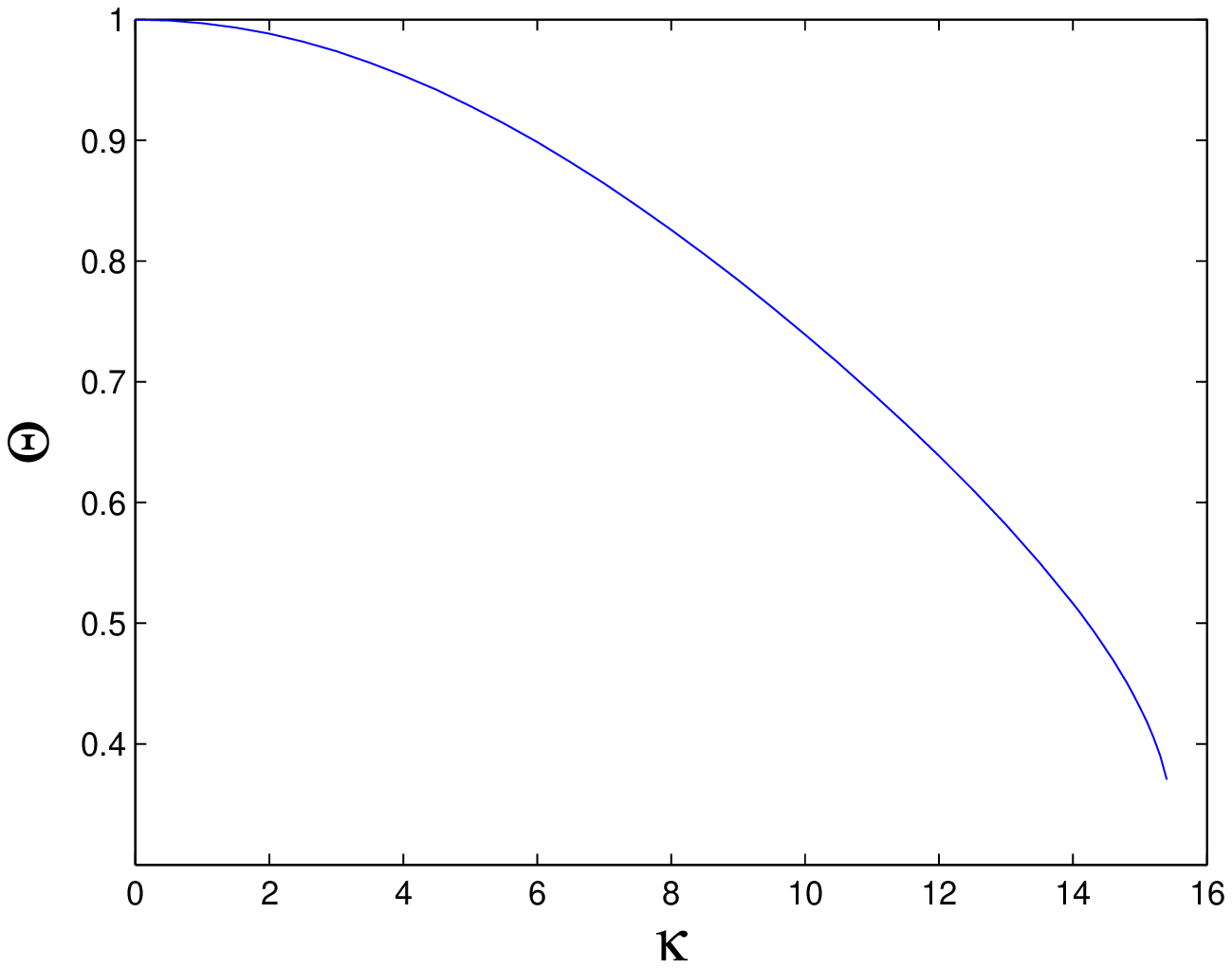}}
\caption{(Color online) The asymmetry ratio, defined as per Eq. (\protect\ref%
{theta}), as a function of the linear-coupling constant in the misaligned ($%
\Delta _{1}=\protect\pi ,\Delta _{2}=0$) attraction-attraction model with
fixed total norm of the two-component soliton, $N=5$. (a) The
pseudo-bifurcation. (b) The termination of the family of asymmetric soliton
solutions.}
\label{offset_AA_bif}
\end{figure}

Further, Fig. \ref{offset_AA_bif} depicts the dependence of the asymmetry
ratio, defined as per Eq. (\ref{theta}), on coupling coefficient $\kappa $.
In the case of the maximum horizontal mismatch, $\Delta _{1}=\pi ,\Delta
_{2}=0$, we observe that the bifurcation which breaks the (quasi-)symmetry
of solitons at a critical value of $\kappa $ is replaced by the \textit{%
pseudo-bifurcation}, i.e., the branch of asymmetric solutions asymptotically
approaches its quasi-symmetric counterpart with the increase of $\kappa $,
but the bifurcation does not happen, since the two branches never merge. A
similar phenomenon was reported in the 1D model with $\Delta =\pi $ \cite{we}%
. The replacement of the bifurcation by the pseudo-bifurcation is observed
only in the case of the largest mismatch, $\Delta _{1}=\pi $ and/or $\Delta
_{2}=\pi $, and it explains the above-mentioned total instability of the
family of quasi-symmetric solitons, as the corresponding branch never has a
chance to get stabilized by the inverse bifurcation (if one considers the
evolution of the solutions with the increase of $\kappa $ at fixed total
norm $N$).

A new feature specific to the 2D system is that, in the system with the
largest diagonal mismatch, $\Delta _{1}=\Delta _{2}=\pi $, the numerical
procedure suddenly ceases to converge at some critical value of $\kappa $,
and no asymmetric solitons are found past this point. For instance, at $%
\kappa =15.485$ a stable asymmetric soliton with a regular profile can be
found, but when the coupling increases to $\kappa =15.490$, the soliton is
lost. The shape of the $\Theta (\kappa )$ dependence in Fig. \ref%
{offset_AA_bif}(b) strongly suggests that the branch of asymmetric solitons
terminates, in this case, due to a tangent bifurcation, which results from
collision and mutual annihilation of the branch with an additional branch of
unstable solutions, that we did not aim to find.

\begin{figure}[tbp]
\includegraphics[width=3in]{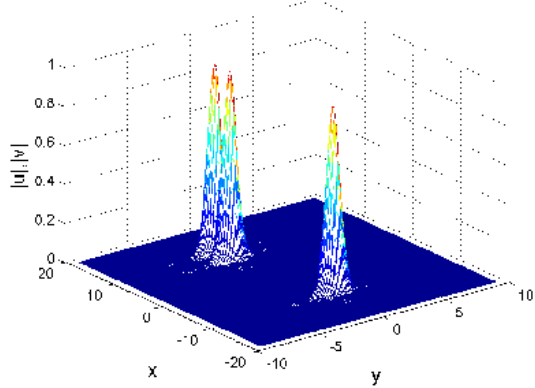}
\caption{(Color online) A stable soliton in the diagonally mismatched ($%
\Delta _{1}=\Delta _{2}=\protect\pi $) attraction-attraction system. In this
case $\protect\mu =-9$, and $N_{u}=0.8$, $N_{v}=1.6$.}
\label{offset_AA_2x1}
\end{figure}

Other types of stable solitons and vortices have also been found in the
misaligned system of the AA type. In particular, an example of a soliton
with two peaks in one component and a single peak in the other is shown in
Fig. \ref{offset_AA_2x1}.

\begin{figure}[tbp]
\includegraphics[width=3in]{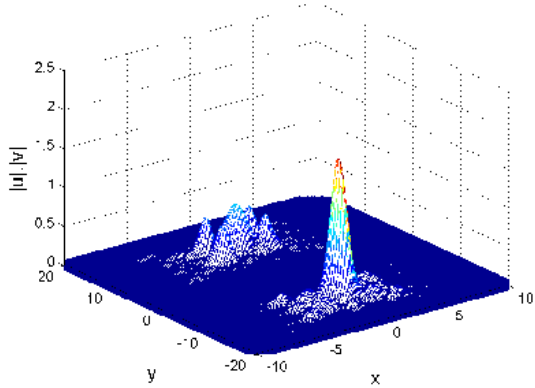}
\caption{(Color online) An example of a strongly asymmetric stable soliton
in the diagonally mismatched ($\Delta _{1}=\Delta _{2}=\protect\pi $)
repulsion-repulsion system with $\protect\kappa =10$. This soliton has $%
\protect\mu =-11$, and $N_{u}=4.4$, $N_{v}=1.6$.}
\label{offset_RR_asym}
\end{figure}

Stable asymmetric solitons in the RR system with misaligned lattices have
been found too. A typical stable asymmetric soliton in this system is shown
in Fig. \ref{offset_RR_asym}

\section{Conclusion}

We have introduced experimentally relevant models of two stacked flat BEC\
traps, with attractive or repulsive interactions between atoms. Each flat
trap carries an OL (optical lattice), stable 2D solitons being impossible
without it. First, it was demonstrated that the linear coupling splits every
finite bandgap in the linear spectrum of the single-component model into two
subgaps, which are separated by narrow Bloch bands. The main issue,
addressed in this work by means of the VA (variational approximation) and
numerical methods, is SSB\ (spontaneous symmetry breaking) in families of
solitons and localized vortices, as a result of the competition of three
factors: the attractive or repulsive nonlinearity in each core, the action
of the OL potential, and the linear coupling between the cores. Similar to
what was found in the zero-dimensional setting \cite{double-well} (this
means a double-well potential without any transverse dimension) and
one-dimensional models \cite{we,Michal}, the SSB occurs in families of
symmetric or antisymmetric states, in the case of the attractive or
repulsive nonlinearity, respectively. In either case, the corresponding
bifurcation destabilizes the branch of symmetric or antisymmetric solitons
or vortices, giving rise to a stable branch of asymmetric solutions. The VA,
based on the Gaussian ansatz, yields an accurate prediction for the
bifurcation and the branch of asymmetric solitons generated by the
bifurcation. A feature specific to the 2D setting is the termination of all
stable branches in the AA (attraction-attraction) system due to the onset of
collapse.

Stable asymmetric solitons sitting in higher finite bandgaps, and localized
vortices with multiple values of the topological charge have been found too.
In addition, the models considered in this work give rise to first examples
of (stable) embedded solitons and embedded vortices\textit{\ }(localized
states found inside Bloch bands separating the subgaps) in any 2D setting.

Solitons and localized vortices were also considered in the linearly-coupled
systems whose symmetry is broken by opposite signs of the nonlinearity in
the two traps, or by a mismatch between the OLs in them. In the former case,
the system gives rise to two distinct types of stable solitons and vortices,
which are dominated by either a self-attractive component or self-repulsive
one, which sit, respectively, in the semi-infinite gap, or in a finite
bandgap. In the system with mismatched lattices, the phenomenon of the
pseudo-bifurcation (which was recently reported in the 1D system \cite{we})
was found: when the mismatch takes the largest value ($\pi $) in any
direction (horizontal or diagonal), the SSB bifurcation fails to happen, as
branches of asymmetric and quasi-symmetric (or quasi-antisymmetric)
solutions asymptotically approach each other, but never merge.

This work was supported, in a part, by the Israel Science Foundation through
the Center-of- Excellence Grant No. 8006/03 and the German-Israel Foundation
(GIF) through Grant No. 149/2006.

\end{document}